\documentclass[11pt,twocolumn]{article}
\usepackage{xcolor}
\usepackage[normalem]{ulem}   %

\usepackage{amsmath}    %
\usepackage{amsfonts}   %
\usepackage{graphicx}   %
\usepackage[super,articletitle=true]{achemso}
\usepackage{hyperref}   %
\usepackage{geometry}   %
\usepackage{titlesec}   %
\usepackage{fancyhdr}   %
\usepackage{xcolor}
\usepackage{subcaption}
\usepackage{hyperref}
\usepackage{xr-hyper}
\usepackage{soul}

\usepackage[font=small]{caption}
\makeatletter
\renewcommand{\thefootnote}

\makeatother

\titleformat{\section}{\large\bfseries}{\thesection.}{1em}{}
\titleformat{\subsection}{\normalsize\bfseries}{\thesubsection.}{1em}{}

\title{\Large \textbf{Data-driven Design of Metal-Organic Frameworks with Tunable Negative Thermal Expansion}}

\author{
    \begin{minipage}[t]{\textwidth}  
    \centering
    {Prathami Divakar Kamath}\textsuperscript{1,2}, {Francesco Tavani}\textsuperscript{3,4,5}, Alin Marin Elena\textsuperscript{6}, Théo Jaffrelot Inizan\textsuperscript{2,3,4}, Yen-hsu Lin\textsuperscript{3}, Jian Yin\textsuperscript{3}, Wenqian Xu\textsuperscript{7}, Omar M. Yaghi\textsuperscript{3,\color{blue}†},  Kristin A. Persson\textsuperscript{1,2,\color{blue}†}\\
    \vspace{0.5cm}
    \textsuperscript{1}Department of Materials Science \& Engineering, University of California, Berkeley, CA, USA\\
    \textsuperscript{2}Materials Sciences Division, Lawrence Berkeley National Laboratory, Berkeley, CA, USA \\
    \textsuperscript{3}Department of Chemistry, University of California, Berkeley, CA, USA \\
    \textsuperscript{4}Bakar Institute of Digital Materials for the Planet, Division of Computing, Data Science, and Society, University of California, Berkeley, CA, USA \\ 
    \textsuperscript{5}Dipartimento di Chimica, Università degli Studi di Roma La Sapienza,\\ P.le A. Moro 5, I-00185, Rome, Italy \\
    \textsuperscript{6}Scientific Computing Department, Science and Technology Facilities Council, Daresbury Laboratory, UK\\   
    \textsuperscript{7}X-ray Science Division, Advanced Photon Source, Argonne National Laboratory, Lemont, IL, USA
\\
        \vspace{0.5cm}
    \textsuperscript{\color{blue}†} 
    Corresponding author:  yaghi@berkeley.edu, kristinpersson@berkeley.edu
    \end{minipage}}
\date{} %
\newcommand{\codename}[1]{{\sc {#1}}}
\newcommand{\abins}{\codename{abins}}
\newcommand{\mantid}{\codename{mantid}}

\begin{document}

\twocolumn[
\maketitle
\begin{center}
    \large \textbf{Abstract}
\end{center}
\begin{center}
    \begin{minipage}{0.9\textwidth}
        \normalsize

Materials with negative thermal expansion (NTE) are essential for applications requiring precise control of thermal expansion. Owing to their exceptional chemical tunability, flexible architectures, and low-energy lattice vibrations, metal–organic frameworks (MOFs) represent a rich platform for exploring NTE. However, uncovering the structural motifs that govern NTE across the enormous MOF design space remains experimentally challenging, and large-scale first-principles phonon calculations are computationally prohibitive. Here, we comprehensively evaluate the factors influencing NTE in MOFs by utilizing a high-throughput workflow based on MACE-MP-MOF0, a machine learning interatomic potential fine-tuned for MOFs with near-ab initio accuracy, to construct PhononMOFdb, a database of phonons, inelastic neutron scattering spectra, bulk moduli, and heat capacities for over 12,000 MOFs. High-throughput screening of a representative subset of 7,000 MOFs from this database reveals that highly porous cubic topology frameworks with heavier, lower-valent metal nodes favor strong NTE, while linker functionalization provides a practical handle for tuning NTE magnitude and sign without compromising mechanical stability. Experimental validation via high-resolution temperature-dependent synchrotron powder X-ray diffraction on the Ce-UiO-66 MOF and its brominated variants confirms the design recipe. This work establishes a data-driven strategy for engineering NTE in MOFs, showing how machine learning-accelerated discovery and targeted experimental validation together unlock predictive materials design.
  \end{minipage}
\end{center}
]

\section{Introduction}

 Negative Thermal Expansion (NTE) \cite{nte1,nte2,nte3} materials contract upon heating and are of significant interest because they enable precise control of thermal expansion in composites for applications including precision optics, microelectronics, and aerospace engineering \cite{nteapp,nteapp1,nteapp2,nteapp3,nteapp4}. While NTE is uncommon in conventional solids, it is widely observed in metal-organic frameworks (MOFs) \cite{wu2008negative,isoreticular,zhou2008origin}, which are ultra-porous materials with exceptional structural tunability and flexibility \cite{mof,mof1,mof2}. The flexible nature of the MOF structures gives rise to a high density of low-frequency phonons \cite{thermal,D3TA02214E}, which are collective atomic vibrations that govern how a crystal responds to heat \cite{phonons, Ziman2001}, making MOFs ideal candidates for tuning NTE. Enhancing NTE is desirable for designing zero thermal expansion composites,\cite{takenaka_negative_2012} whereas suppressing NTE can mitigate excessive thermal contraction and improve MOF stability \cite{evans2019assessing}. Despite the widespread occurrence of NTE in MOFs, achieving systematic and predictive control over its magnitude and sign remains challenging due to the large chemical space for exploration and expensive phonon computations \cite{kamath,kamencek2022understanding}.

Although experimental strategies based on local node distortion \cite{transition,burtch2019negative}, mixed-linker solid solutions \cite{baxter2019tuning}, defect engineering \cite{defect}, and framework retrofitting \cite{retrofitting} have successfully tuned NTE in MOFs, these approaches have primarily been demonstrated within specific material families and are not yet readily generalizable across the broader MOF chemical space. Furthermore, NTE measurements for a given MOF may vary with sample crystallinity, defect density, and thermal history \cite{vornholt2024node}, making it difficult to benchmark  quantitative models. High-throughput computational screening \cite{yue2024classifying} and targeted mechanistic studies \cite{burtch2019negative,evans2019assessing,isoreticular} have established that NTE arises predominantly from low-energy transverse linker phonons and rigid-unit modes (RUMs) \cite{rums,kamencek2022understanding}, yet these theoretical investigations relied on classical force fields \cite{uff,mofff} known to be affected by larger errors in the modeling of MOF phonons compared to density functional theory (DFT) \cite{Wieser2024,vmof}. Further, existing DFT phonon datasets \cite{Moosavi,yue} cover only a small fraction of the MOF chemical space, and the approximations required for large-scale calculations frequently introduce spurious imaginary modes that corrupt derived thermal properties \cite{spurious}. Machine learning interatomic potentials \cite{loew2025universal,mofsim} now bridge the computational gap, with MACE-MP-MOF0 \cite{kamath} delivering ab initio-quality phonons and derived thermal and mechanical properties at a fraction of the DFT cost. Nevertheless, a transferable, phonon-centric design methodology for predictively tuning NTE across a broad MOF chemical space remains absent.

In this work, we leverage MACE-MP-MOF0 (a fine-tuned MACE-MP-0 \cite{mace} model on MOF structures for accurate phonon predictions) to construct PhononMOFdb, the largest MOF phonon database to date, comprising $\sim$12,000 structures with phonon densities of states (DOS), band structures, heat capacities, and derived vibrational thermodynamic quantities, along with bulk moduli and inelastic neutron scattering (INS) spectra validated against DFT and experiments. This database enables a vibration-based screening strategy that identifies the low-frequency modes governing NTE and reveals that heavy metal nodes in cubic, high-porosity topologies amplify their contributions. We show that systematic linker functionalization continuously tunes thermal expansion, including negative to positive transitions, without altering original framework stability. We validate these design principles with high-resolution synchrotron powder X-ray diffraction (PXRD) experiments on Ce-UiO-66 and two of its brominated variants that confirm our predictions of decreased NTE with increased bromine loading due to suppressed RUMs and transverse vibrations. Altogether, this work establishes a data-driven, phonon-centric framework for tuning thermal expansion across diverse MOFs, demonstrating how computational predictions can guide experimental realization of targeted NTE behavior.

\section{Materials and Methods}
\begin{figure*}[t]
    \centering
    \includegraphics[width=\linewidth]{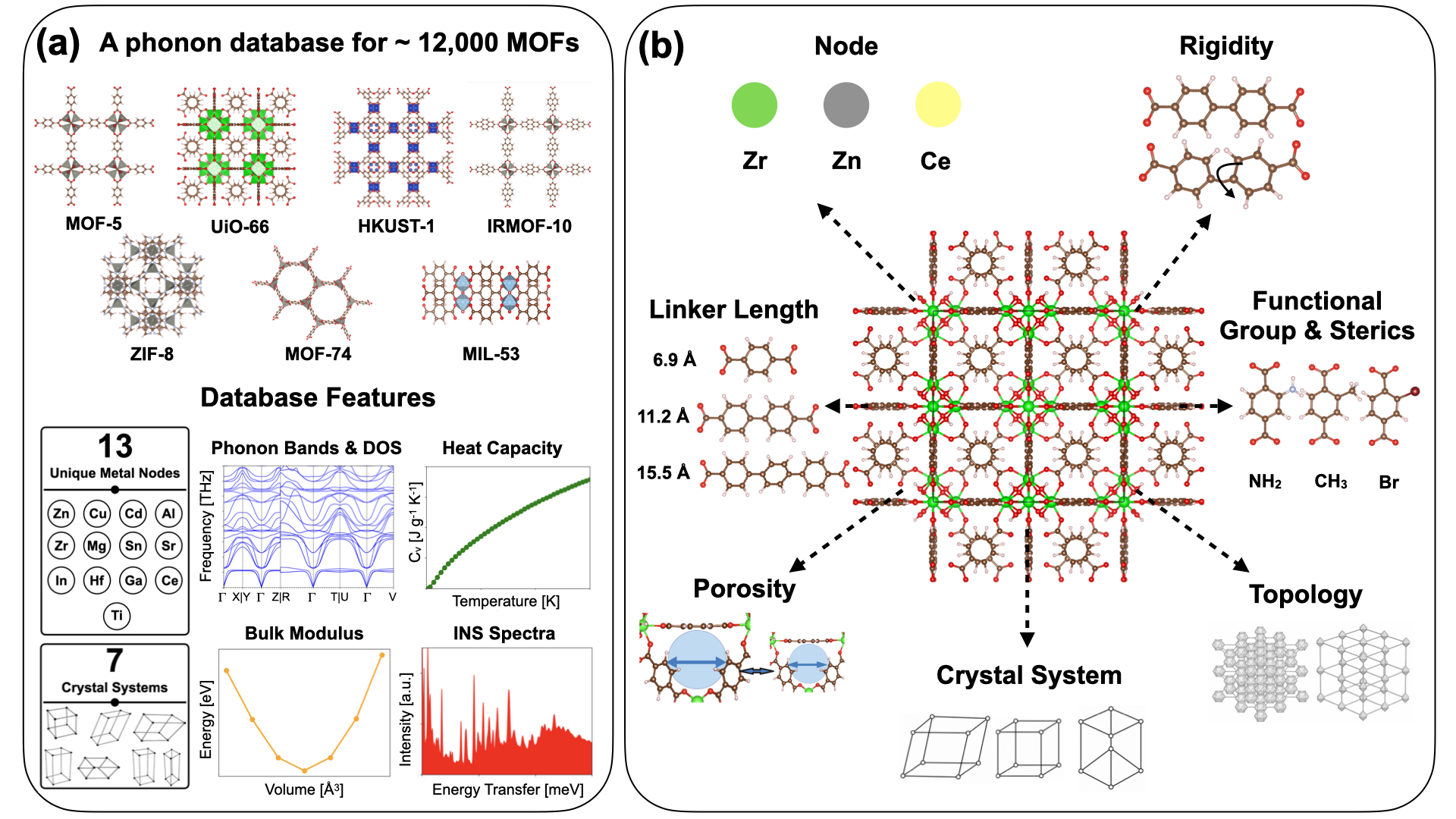}
    \caption{(a) Overview of the PhononMOFdb database, consisting of phonons and thermophysical properti es for over 12,000 MOFs. All seven crystal systems are covered with the following distribution: 46.9\% triclinic, 37.5\% monoclinic, 8.2\% orthorhombic, 2.8\% trigonal, 2.7\% tetragonal, 1.3\% hexagonal and 0.6\% cubic (b) Schematic representation of the data-driven screening workflow: seven tunable structural descriptors are systematically analyzed against the low-frequency vibrational landscape to identify colossal NTE candidates.}
    \label{fig: figure1}
\end{figure*}
\subsection{PhononMOFdb Generation}
PhononMOFdb was generated by applying the MACE-MP-MOF0 (v2) model\cite{kamath} to $\sim$12,000 computation-ready MOFs from the QMOF database\cite{qmof1,qmof2} because it provides a valuable collection of computation-ready MOFs with the highest structural fidelity, exhibiting 10$\times$ lower error rates than other similar repositories \cite{white_high_2025,chung_advances_2019}. This database was defined by retaining every QMOF structure built exclusively from the 24 elements represented in the MACE-MP-MOF0 fine-tuning set, which corresponds to $\sim$60\% of the QMOF database and results in the thirteen metals in nodes captured in Figure \ref{fig: figure1}. The composition of the training set, the resulting domain of applicability, and the known limitations of this MLIP relevant to this screening are summarized in Section S2 of the SI. Harmonic phonon spectra and derived thermal properties were computed using Phonopy \cite{phonopy-phono3py-JPCM,phonopy-phono3py-JPSJ}. More than 78\% of the structures are dynamically stable, with only $\sim$11\% exhibiting an imaginary-mode fraction exceeding 0.1\%. In addition, bulk moduli were obtained via the Birch–Murnaghan equation of state using seven volumes spanning $\pm$3\% linear strain about the equilibrium volume. INS spectra were simulated at 10 K using \abins{} within \mantid,\cite{Dymkowski2018AbINS,arnoldMantidDataAnalysis2014} employing a TOSCA-geometry incoherent approximation including two-phonon transitions \cite{Pinna2018NeutronGuideTOSCA}. Additional details are provided in SI Section S1.

\subsection{DFT phonon computations}
The DFT INS data to benchmark PhononMOFdb accuracy was generated for ZIF-8 from harmonic phonons computed with the VASP 6.4.2 package \cite{vasp,vasp1,vasp2,vasp3}. The PBE density functional \cite{pbegga} with D3(BJ) \cite{bj,d3} empirical dispersion correction was applied to account for long-range van der Waals interactions. The energy convergence was set to $10^{-8}$ eV for the self-consistent field (SCF) cycles, and the force convergence was set to -0.02 \text{eV/\AA} for a tight structural relaxation. A $2 \times 2\times2$ supercell containing 1104 atoms was used on the primitive cell for finite difference phonon calculations with Phonopy (SI Section S7).

\subsection{MOF synthesis and synchrotron PXRD characterization}
The Ce-UiO-66-X (X = H, Br, 2Br) MOFs were synthesized following a previously reported solvothermal procedure, using dimethylformamide as solvent \cite{GAO2024124597}. The MOF phase purity and crystallinity were verified by laboratory-based PXRD measurements. Synchrotron PXRD measurements were performed using a wavelength of 0.2099 \text{\AA} on beamline 11-ID-B at the Advanced Photon Source (APS), Argonne National Laboratory. Calibration of the PXRD patterns was performed with CeO$_2$ powder as a standard, and GSAS-II \cite{GSASII} was employed for azimuthal integration. The samples were washed with ethanol and dried under reduced pressure at room temperature, as thermal activation has been reported to potentially cause framework collapse in Ce-UiO-66-X analogues \cite{son2020solvent,gonzalez2024adsorption}. Thermogravimetric analysis (TGA) shows that residual ethanol remains in all three frameworks and is lost from room temperature upward (SI Section S4), so the measured samples are partially solvated, in contrast to the guest-free frameworks modeled computationally. Consistent with this, Ce-UiO-66-Br shows peak broadening, loss of diffraction contrast and an abrupt lattice contraction above 250K. We therefore evaluate the volumetric thermal expansion coefficients $\alpha_V$ only over the common 100–250K window, where the lattice of all three frameworks remains cubic with no new reflections. Lattice parameters were extracted at each temperature by Le Bail refinement in GSAS-II (Figure S1), with $wR \leq 4.5\%$ for all fits (Tables S1–S3). 
\section{Results and Discussion}
\begin{figure*}[t]
    \centering
    \includegraphics[width=\linewidth]{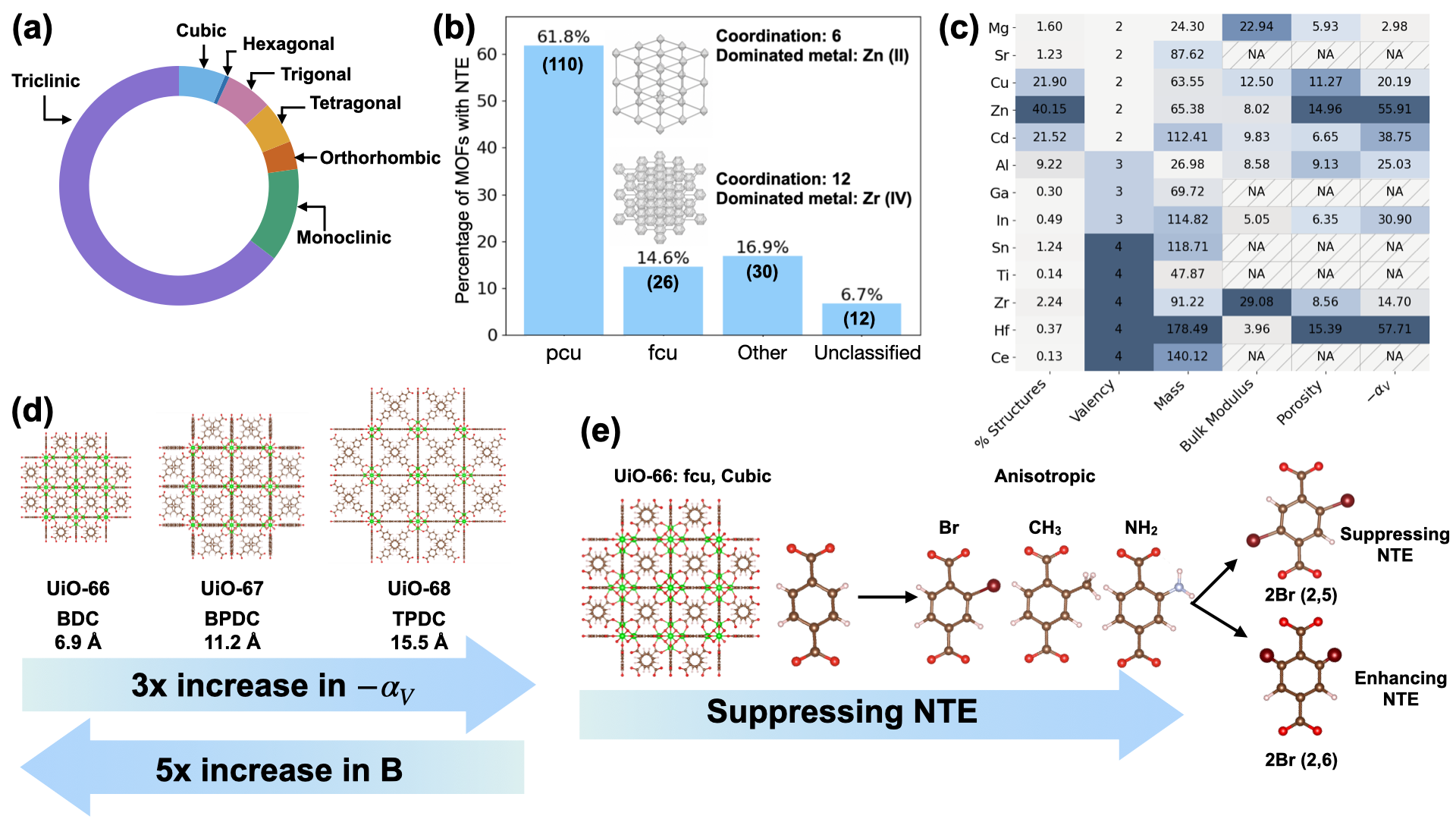}
    \caption{(a) Distribution of volumetric NTE candidates across all seven crystal systems (43.9\% triclinic, 40.7\% monoclinic, 8.6\% orthorhombic, 2.8\% trigonal, 2.5\% tetragonal,  1.0\% hexagonal, 0.5\% cubic). (b) Distribution of NTE candidates by topology where "Unclassified" denotes structures with unassigned topologies. (c) Metal-node descriptor matrix where colors are normalized independently for each column, with darker shades indicating larger values. The \% structures correspond to the elemental distribution in the initial screening pool. (d) Effect of isoreticular linker elongation on NTE and bulk modulus. (e) Linker functionalization enables systematic tuning of NTE magnitude and sign. The corresponding NTE values for each structure in (d) and (e) are listed in Tables S8–S9.}
    \label{fig: figure2}
\end{figure*}

\subsection{Database Generation and High-Throughput Screening}
As illustrated in Figure \ref{fig: figure1}a, PhononMOFdb covers all seven crystal systems, moving beyond previous work mainly focused on isoreticular cubic MOFs. \cite{isoreticular} The database includes thirteen unique metals in nodes with diverse linkers and functional groups which enables an extensive study of how structural factors influence the vibrational landscape of MOFs. The heat capacities of MOFs in PhononMOFdb can be used to estimate the energy penalty associated with adsorbent retrieval \cite{Moosavi,spurious} while the bulk moduli provide an estimate for the MOF mechanical stability \cite{SNURR201926}. The reliability of MACE-MP-MOF0 for predicting the aforementioned properties relative to experiments has been demonstrated in previous studies \cite{kamath,spurious,mofsim}. Here, we validate the accuracy of INS spectra in PhononMOFdb by benchmarking against both DFT and experimental measurements. 

In MOFs, inelastic neutron scattering (INS) has been widely used to characterize framework lattice dynamics, gas adsorption, and adsorbate-induced changes in vibrational spectra \cite{bahadur2025competitive, casco2016paving, callear2013high, chen2023analysis}. MACE-MP-MOF0 was recently evaluated against experimental INS spectra in the uMOF benchmark \cite{umof} for six well-known MOFs spanning different symmetries, metal nodes and linkers, including MOF-5(Zn), MIL-53(Al), MIL-140A(Zr), ZIF-8(Zn), ZIF-4(Zn) and ZIF-7(Zn), where it ranked among the most accurate of five MLIPs with modest red shift of $\leq$ 1.8 meV relative to experiments which is a well-known consequence of the PBE functional \cite{under,under1} (Figure S4 and Table S4) and was the only one to yield 0\% imaginary modes. Since broadly accessible experimental INS data for MOFs remain limited \cite{cheng2023database}, PhononMOFdb provides an open-access resource for INS spectra with a released workflow to enable extension to additional MOFs beyond the current dataset.

While isotropic MOFs are capable of exhibiting large NTE \cite{isoreticular}, the vast diversity of MOFs also includes anisotropic frameworks that can exhibit linear or areal NTE, in which two principal axes contract while the third expands showing highly anisotropic NTE, while maintaining an overall non-negative volumetric expansion coefficient \cite{massive}. Although less explored, the mechanism for observing NTE in anisotropic MOFs has been attributed to a combination of low-frequency transverse linker vibrations and RUMs \cite{ortho}, predominantly in the 0–2 THz range \cite{kamencek2022understanding}. This is consistent with the previously identified mechanisms in isoreticular MOFs \cite{zhou2008origin, isoreticular} suggesting a common phonon-driven origin of NTE across diverse MOF families. 

Motivated by these observations, we leverage the low-frequency (0–2 THz) harmonic phonon data in PhononMOFdb to develop a computationally efficient screening strategy for volumetric NTE. We build on this correlation and screen a representative subset of 7,000 MOFs from PhononMOFdb (Table S5) for intense low-frequency phonon modes, identifying 432 candidates for quasi-harmonic approximation (QHA) calculations. As shown in Figure S5b, the QHA calculations reproduce the experimental NTE trends across several well-characterized MOFs. Of the 432 candidates, we retain 390 MOFs by excluding MOFs exhibiting imaginary modes above 0.1\% in their QHA calculations as such dynamical instabilities are common in MOFs at compressed volumes and are typically not accounted for in high-throughput screening studies. Finally, to address the challenge of achieving large NTE without softening the framework to the point of mechanical failure \cite{evans2019assessing,linnera2019negative}, we apply a bulk modulus threshold of $B \geq 1$ GPa at 100 K, reducing the pool to 178 mechanically stable volumetric NTE candidates.

To investigate the molecular-level origins of NTE, we then sequentially analyze the dependence of the NTE coefficient $\alpha_V$ on seven structural descriptors for the diverse screened frameworks, including MOF crystal system, topology, metal node identity, linker length, flexibility and functionalization (see Figure \ref{fig: figure1}b).
 Given that the QMOF database, and hence PhononMOFdb, exhibits an uneven distribution across the structural descriptors (Figure S1), raw counts of screened candidates are biased toward abundant categories, partly reflecting the composition of the screening pool rather than an intrinsic preference for NTE. We therefore define an enrichment ratio for each descriptor category as the fraction of screened NTE candidates belonging to that category divided by its fraction in the 7k screening subset, so that values above 1 directly identify the structural features that favor volumetric NTE, independent of how common they are in the database.

\subsubsection{Step 1: Role of Crystal System}
Figure \ref{fig: figure2}a shows the crystal system distribution of the 178 screened MOFs with volumetric NTE. Colossal volumetric NTE (defined as $|\alpha_V| \geq 100$) is not confined to cubic frameworks, which dominate the experimental literature \cite{ortho}, but extends across all seven crystal systems, with triclinic and monoclinic MOFs exhibiting the largest NTE magnitudes (Figure S6a), largely reflecting their prevalence in the anisotropic QMOF database (Figure S6b). The enrichment ratios (Figure S7a) separate this abundance effect from structural preference where cubic frameworks are enriched tenfold and tetragonal and trigonal frameworks twofold, consistent with high-symmetry lattices favoring volumetric NTE. Whereas low-symmetry frameworks are depleted. Triclinic frameworks are the notable exception, being mildly enriched (1.4$\times$) and forming the majority of candidates. This apparent contradiction arises because small geometric perturbations from linker functionalization or distortions introduced during geometry relaxation, lower the space-group symmetry without altering the underlying node–linker connectivity. Eleven MOFs emerge as candidates for colossal volumetric NTE, spanning triclinic, monoclinic, trigonal, tetragonal and cubic systems highlighting the untapped potential of anisotropic MOFs for extreme volumetric NTE. This suggests that crystal system alone does not explain large NTE and motivates a search for common governing structural features.

\subsubsection{Step 2: Topology and Metal Node Effects}
We next analyze MOF topology, which encodes node–linker connectivity and is invariant to small functionalization- or relaxation-induced distortions, thereby providing a more robust structural descriptor than crystal system. As shown in Figure \ref{fig: figure2}b, 61.8\% of screened MOFs adopt the primitive cubic (\textbf{pcu}) topology, followed by face-centered cubic (\textbf{fcu}). Notably, seven of the eleven colossal NTE MOFs are also \textbf{pcu} frameworks despite differing crystal systems, highlighting a clear structure–property correlation. The enrichment ratios (Figure S7b) confirm this preference where \textbf{pcu} (4.1$\times$) and \textbf{fcu} (8.1$\times$)  are strongly enriched as their well-defined three-dimensional net with cubic-like structural motifs favor volumetric NTE. Interestingly, 95\% of triclinic NTE MOFs retain cubic topologies despite their \textit{P1} space group, highlighting cubic connectivity rather than crystal system as the relevant feature for NTE.

The screened \textbf{pcu} frameworks are almost exclusively Zn-based, consistent with the preference of Zn$^{2+}$-based nodes for sixfold coordination, whereas all \textbf{fcu} MOFs contain Zr$^{4+}$-based nodes that form stable 12-coordinate clusters. Topology is a stronger predictor of NTE here than metal node alone since \textbf{pcu} frameworks are enriched among NTE candidates 2$\times$ more than Zn-based frameworks (Figure S7b, Table S5), as Zn also forms many non-\textbf{pcu} frameworks that didn't exhibit NTE. On average, we find that the large majority of \textbf{pcu} MOFs follows from large uniform pores and soft frameworks enabled by this topology (see Table S6) which favor large transverse vibrations and enhance NTE. The \textbf{fcu} MOFs exhibit higher bulk moduli and lower volumetric NTE than \textbf{pcu} frameworks as a result of their significantly higher metal-linker coordination and lower porosity that stiffens the framework. This trend is corroborated by the two best-characterized volumetric NTE MOFs in the experimental literature, namely MOF-5 (\textbf{\textbf{Zn-pcu}}) \cite{insmof-5} and UiO-66 (\textbf{\textbf{Zr-fcu}}) \cite{burtch2019negative}, which exhibit large and moderate NTE, respectively. The \textbf{\textbf{fcu}} topology therefore represents a viable design target when mechanical stability is a priority.

Figure \ref{fig: figure2}c displays the metal node distribution of the MOFs selected from the screening. Table S5 indicates whether an element is over or underrepresented among NTE MOFs independent of its abundance. Five metals (i.e., Sr, Ga, Sn, Ti and Ce) yield no screened candidates and are labeled "NA", reflecting the limited screening pool rather than intrinsically unfavorable NTE. The remaining 178 candidates span eight distinct metal nodes. Zr and Hf-based frameworks are the most strongly enriched (6.5 and 3.0) $\times$, respectively in Table S5), followed by Zn with an almost two fold enrichment, mirroring the dominance of the \textbf{fcu} and \textbf{pcu} topologies discussed above. Conversely, Cd, Cu and Al-based frameworks are depleted. We interpret the depletion of Cu with caution, given the elevated model errors for spin-polarized nodes and the known experimental NTE of HKUST-1 \cite{wu2008negative}. Among the screened candidates, the largest average NTE magnitudes are found in Zn and Hf-based MOFs, consistent with their larger average pore volumes in the descriptor matrix of Figure \ref{fig: figure2}c and the established correlation between porosity and NTE \cite{evans2019assessing}. Furthermore, when choosing among metals that satisfy the same topology coordination requirements, a heavier node could further enhance NTE by lowering the frequencies of the NTE-driving modes, particularly at low temperature, as shown by substituting Zr with Ce in UiO-66 (Figure S13; SI Section S9.2).

\subsubsection{Step 3: Linker Effects}
We next explore linker chemistry as an additional tunable handle to enhance NTE for a given MOF topology and metal node. Longer linkers and increased porosity have been shown to generally favor larger NTE in previous studies\cite{evans2019assessing,baxter2019tuning}. As illustrated in Figure \ref{fig: figure2}d, systematic isoreticular expansion of the linkers from 1,4-benzenedicarboxylate (BDC) in UiO-66 to 1,4-Di(4-carboxyphenyl)benzene (TPDC) in UiO-68 leads to an approximately three-fold increase in NTE but simultaneously reduces the bulk modulus by nearly a factor of five. Based on these findings, organic linker elongation is a viable route for enhancing NTE in MOFs with exceptionally high stiffness such as UiO-66 ($\sim$35 GPa). However, this strategy is unlikely to be transferable to MOFs with intrinsically lower bulk moduli, where further mechanical softening would compromise structural integrity (see Figure S5 and Table S7).

We then analyze linker functionalization as a tool to modify the underlying lattice dynamics and NTE behavior without compromising the mechanical stability of the core framework. The screened \textbf{\textbf{pcu}} and \textbf{\textbf{fcu}} MOFs (see Figure S10) predominantly feature BDC-derived linkers incorporating bulky side groups, extended conjugated backbones, as well as electron-donating (-OH, -NH$_2$ and -CH$_3$) and electron-withdrawing substituents (-F, -Cl, -Br).

\begin{figure*}[t]
    \centering
    \includegraphics[width=\linewidth]{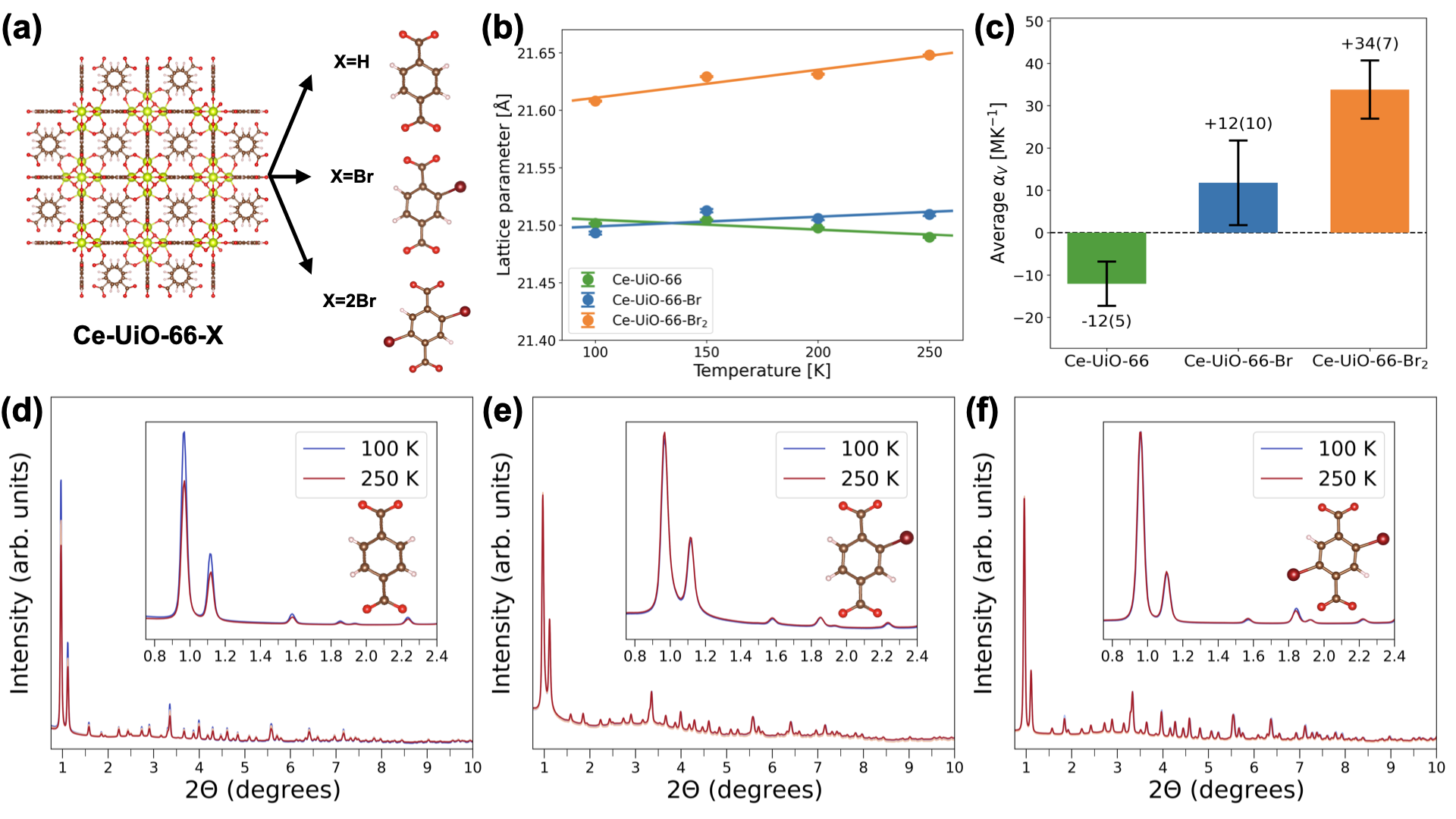}
    \caption{(a) Crystal structure of Ce-UiO-66, highlighting the molecular structures of its three BDC-derived linker variants: pristine (H), mono-brominated (Br), and di-brominated (2Br). (b) Le Bail-refined lattice parameters as a function of temperature for Ce-UiO-66-X (X = H, Br, 2Br). (c) Average volumetric thermal expansion coefficients $\alpha_V$ obtained from the slopes of the fits in (b), showing progressive suppression of NTE with increasing bromination; error bars are the standard errors of the slopes. (d–f) Synchrotron PXRD patterns collected between 100 and 250K for Ce-UiO-66-H (d), Ce-UiO-66-Br (e) and Ce-UiO-66-2Br (f). Insets show a magnification of the patterns recorded at 100 and 250K.}
    \label{fig: figure3}
\end{figure*}

To better understand how MOF linker functionalization influences NTE, we substitute the H atom in the 2-position of the BDC linker in the \textbf{fcu} UiO-66 with X= NH$_2$, CH$_3$ and Br as representative electron donating and withdrawing groups. As shown in Figure S7, electron-donating groups increase vibrational rigidity, while electron-withdrawing groups soften the vibrations, in agreement with rotational rigidity trends previously reported for UiO-66-X frameworks \cite{imaging}. Despite these differences, all such substitutions reduce $|\alpha_V|$ if compared to that of pristine UiO-66, and induce a positive thermal expansion (PTE) at increasing temperatures (see Figure S8a and Table S8). Our analysis is in agreement with previous experimental observations of lattice expansion in UiO-66-NH$_2$ and UiO-66-Br \cite{taddei2017mixed}. We attribute this suppression of NTE to two cooperating effects: the bulky Br and hydrogen-bonding NH$_2$ groups dampen transverse linker vibrations, while the reduction in symmetry from cubic to triclinic upon functionalization disrupts the coherent isotropic contractions responsible for the strong volumetric NTE. We performed an analogous analysis for the UiO-67 system using X= NH$_2$, Br, and CH$_3$ as substituents, revealing the same trend of reduced NTE (see Figure S8b) in agreement with previous experimental findings \cite{goodenough2020interplay}. Interestingly, we find the transition from NTE to PTE is less likely to occur in the UiO-67 MOF derivatives if compared to the UiO-66 MOF analogues, with only the UiO-67-NH$_2$ system approaching positive thermal expansion near 200 K.

\begin{figure*}
    \centering
    \includegraphics[width=\linewidth]{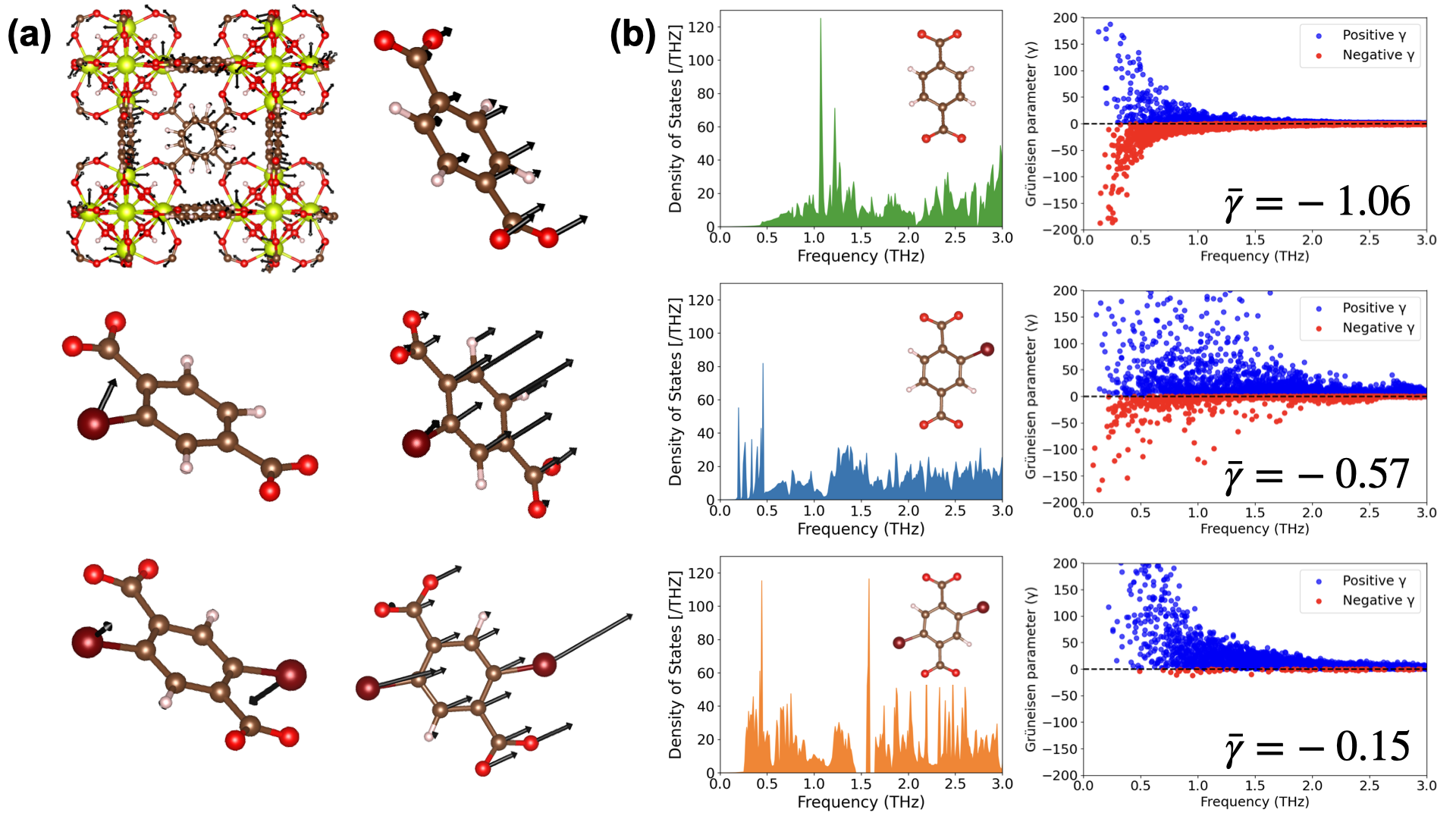}
    \caption{(a) Atomic displacements of the dominant low-frequency phonon modes in the the unit cell of pristine Ce-UiO-66 and BDC (top), 2-bromo-BDC (middle) and 2,5-dibromo-BDC (bottom) organic linkers.  (b) Phonon densities of states (left panels) and Grüneisen parameter ($\gamma$) analysis (right panels) performed for the Ce-UiO-66-X= H, Br, 2Br MOFs.}
    \label{fig: figure4}
\end{figure*}

Having selected Br as a representative heavy substituent whose low-frequency vibrations fall within the 0–2 THz range, we further investigate how increased bromination of the BDC linker affects NTE. As shown in Figure \ref{fig: figure2}e, the introduction of the 2,5-dibromo BDC in the UiO-66 MOF suppresses NTE by limiting transverse vibrations due to a higher degree of steric hindrance of the double-brominated linker if compared to that of 2-bromo BDC. Conversely, incorporating 2,6-dibromo BDC in the UiO-66 MOF activates a complementary transverse vibration mode in which half of the phenyl ring flexes, enhancing linker contractions and producing a larger NTE if compared to that of the single-brominated analogue. 

\subsection{Design Recipe}
Our data-driven analysis suggests a design strategy for the engineering of NTE in MOFs: (i) select a cubic topology, with softer \textbf{pcu} leading to an increased NTE and \textbf{fcu} providing a more rigid starting framework; (ii) within a given topology, choose heavier metal nodes; (iii) favor cubic crystal systems for larger NTE; (iv) tune the length of the linker to balance the NTE and mechanical stability; (v) and functionalize the linker to further modulate the NTE while maintaining the integrity of the MOF framework. 

Taken together, these design principles point toward the \textbf{fcu} topology with a BDC linker as a stable, synthetically accessible starting framework for experimental validation. This connectivity requires a tetravalent metal node, and among available options, the heavier Hf and Ce are favored by the mass–charge arguments established above. Given that NTE has been studied in thermally densified defective Hf-UiO-66 \cite{defect} and pristine Zr-UiO-66 \cite{burtch2019negative}, we select pristine Ce-UiO-66 as a previously uncharacterized candidate predicted to exhibit enhanced NTE. To demonstrate the tunability achievable within a single MOF through linker functionalization, we also investigate the Ce-UiO-66 MOF analogues based on mono- and 2,5 di-brominated BDC variants (see Figure \ref{fig: figure3}a). Computational simulations based on the established design recipe predict Ce-UiO-66 to exhibit larger NTE than Zr-UiO-66, with an increase of linker bromination leading to a progressive NTE suppression (see Figure S9).

\section{Experimental Validation of Design Recipe} 
We experimentally validate the developed data-driven recipe for tuning NTE, focusing on the Ce-UiO-66-X= H, Br and 2Br MOFs. Figures \ref{fig: figure3}b,c display the temperature-dependent evolution of the unit-cell lattice and the resulting average $\alpha_V$ over 100–250K obtained for Ce-UiO-66-X = H, Br, 2Br from synchrotron in situ PXRD measurements \cite{MOF1f, MOF2f,MOF3_f,catal13101338} (Figures \ref{fig: figure3}d–f). As the measured samples retain residual ethanol (see Methods), whereas the calculations describe guest-free frameworks, we use the experiments to test the predicted trend rather than to compare absolute values. Over this range, the average $\alpha_V$ increases from negative for Ce-UiO-66-H to positive for Ce-UiO-66-Br and Ce-UiO-66-2Br, with Ce-UiO-66-2Br showing the largest value (Figure \ref{fig: figure3}c). Linker bromination therefore progressively suppresses NTE, consistent with the trend predicted by the data-driven design recipe.

To further understand the experimentally observed behavior, we performed a phonon mode analysis of the investigated Ce-UiO-66 MOFs (see Figure \ref{fig: figure4}). The mode Grüneisen parameter ($\gamma$) quantifies the volumetric response per vibrational mode \cite{phonons}, with more negative values indicating stronger contraction as vibrational frequency increases with temperature. All three MOFs display intense DOS features in the 0–2 THz range, but their mode-wise contributions evolve systematically with bromination. As shown in Figure 4b, an increasing number of low-frequency modes acquire positive $\gamma$ as a function of Br content, offsetting the negative contributions and reducing the magnitude of the weighted average $\bar{\gamma}$. Visualization of the corresponding phonon eigenvectors in Figure \ref{fig: figure4}a clarifies the microscopic origin of these trends. In pristine Ce-UiO-66-H, the lowest-frequency modes are RUMs involving rigid rotations of the metal-oxide clusters coupled to the out-of-plane transverse vibrations of the BDC linkers. These soft modes carry strongly negative $\gamma$ and give rise to the large NTE observed experimentally (see Figure 4b, top panels). Mono-bromination alters the low-frequency spectrum where the lowest-frequency modes are now dominated by large-amplitude, localized Br oscillations, while the BDC linker and node clusters remain comparatively rigid (Figure 4b, middle panels). Furthermore, the transverse linker vibrations that drove NTE in the pristine Ce-UiO-66-H MOF are now increasingly impeded at higher frequency by the steric bulk of the Br substituent. Di-bromination further increases the suppression of the NTE-driving modes where the lowest-frequency modes reflect combinatorial vibrations of the two Br groups (see Figure 4b, bottom panels), and the additional Br atom further hinders transverse linker displacements at higher frequencies leading to the lowest $\bar{\gamma}$ and weakest NTE among the series. 

We note that NTE in UiO-type frameworks has also been attributed to distortion of the inorganic clusters \cite{vornholt2024node, transition}, a mechanism reported to become significant at elevated temperatures, such as above $\sim$423 K, where a sufficient population of distorted nodes couples to the extended framework \cite{hofmann2026colossal}. Over the 100–250K range studied here, where thermal energy populates modes up to $\sim$5 THz, the low-frequency rigid-unit and transverse linker modes identified in Figure \ref{fig: figure4} provide a rationale for the observed low-temperature NTE, although we do not rule out contributions from other temperature-dependent mechanisms.
Overall, the described atomistic harmonic phonon analysis quantitatively rationalizes the experimentally observed NTE trends and design recipe.

\section{Conclusion}
In summary, we developed and experimentally validated a general strategy for tuning NTE in MOFs through targeted modification of metal node and linker chemistry while preserving framework stability. The resulting design principles identify (i) cubic topologies as favorable structural scaffolds (ii) lower-valent metal nodes as promoters of larger NTE through softer lattice dynamics, with heavier nodes providing a further handle at low temperature, and(iii) linker functionalization as a means to finely modulate thermal expansion. Validation on selected MOFs of the Ce-UiO-66-X (X = H, Br, 2Br) family confirms these trends and fills a gap in the experimental characterization of NTE effects in the Cerium-based UiO-66 MOFs. Beyond this validation, the eleven colossal NTE candidates identified by our screening provide promising targets for future experimental investigation. More broadly, this work shows how phonon-centric screening of databases such as PhononMOFdb can uncover the vibrational origins connecting framework structure to thermal expansion behavior, establishing design rules that guide experimental synthesis and enable the rational engineering of MOFs with targeted thermal expansion.

\section{Data Availability}
The database and codes are available at \textcolor{blue}{https://doi.org/10.6084/m9.figshare.32050410}
\section{Supplementary Information}
The Supplementary information is available as an additional pdf.

\section{Acknowledgments}
This research used computing resources of the National Energy Research Scientific Computing Center (NERSC), a Department of Energy (DOE) Office of Science User Facility and by Science and Technology Facilities Council Scientific Computing Department’s SCARF cluster. P.D.K. acknowledges financial support from U.S. National Science Foundation's "The Quantum Sensing Challenges for Transformational Advances in Quantum Systems" program No. 2326838.  F.T. acknowledges funding from the European Union’s Horizon Europe research and innovation program under the Marie Skłodowska-Curie grant agreement No. 101201413 (AQUAFRAME). A.M.E. was also supported by the Ada Lovelace Centre at the Science and Technology Facilities Council (https://adalovelacecentre.ac.uk/), the Physical Sciences Data Infrastructure (https://psdi.ac.uk; jointly STFC and the University of Southampton) under grants EP/X032663/1 and EP/X032701/1, and EPSRC under grants EP/W026775/1 and EP/V028537/1. T.J.I acknowledges financial support from the Bakar Institute of Digital Materials for the Planet. This research was performed on APS beam time award(s) (DOI: https://doi.org/10.46936/APS-193506/60016502) from the Advanced Photon Source, a U.S. Department of Energy (DOE) Office of Science user facility operated for the DOE Office of Science by Argonne National Laboratory under Contract No. DE-AC02-06CH11357.

\bibliographystyle{achemso}

\begin{mcitethebibliography}{82}
\providecommand*\natexlab[1]{#1}
\providecommand*\mciteSetBstSublistMode[1]{}
\providecommand*\mciteSetBstMaxWidthForm[2]{}
\providecommand*\mciteBstWouldAddEndPuncttrue
  {\def\EndOfBibitem{\unskip.}}
\providecommand*\mciteBstWouldAddEndPunctfalse
  {\let\EndOfBibitem\relax}
\providecommand*\mciteSetBstMidEndSepPunct[3]{}
\providecommand*\mciteSetBstSublistLabelBeginEnd[3]{}
\providecommand*\EndOfBibitem{}
\mciteSetBstSublistMode{f}
\mciteSetBstMaxWidthForm{subitem}{(\alph{mcitesubitemcount})}
\mciteSetBstSublistLabelBeginEnd
  {\mcitemaxwidthsubitemform\space}
  {\relax}
  {\relax}

\bibitem[Miller \latin{et~al.}(2009)Miller, Smith, Mackenzie, and Evans]{nte1}
Miller,~W.; Smith,~C.; Mackenzie,~D.; Evans,~K. Negative thermal expansion: a review. \emph{J. Mater. Sci.} \textbf{2009}, \emph{44}, 5441--5451\relax
\mciteBstWouldAddEndPuncttrue
\mciteSetBstMidEndSepPunct{\mcitedefaultmidpunct}
{\mcitedefaultendpunct}{\mcitedefaultseppunct}\relax
\EndOfBibitem
\bibitem[Liu \latin{et~al.}(2018)Liu, Gao, Chen, Deng, Lin, and Xing]{nte2}
Liu,~Z.; Gao,~Q.; Chen,~J.; Deng,~J.; Lin,~K.; Xing,~X. Negative thermal expansion in molecular materials. \emph{Chem. Commun.} \textbf{2018}, \emph{54}, 5164--5176\relax
\mciteBstWouldAddEndPuncttrue
\mciteSetBstMidEndSepPunct{\mcitedefaultmidpunct}
{\mcitedefaultendpunct}{\mcitedefaultseppunct}\relax
\EndOfBibitem
\bibitem[Chen \latin{et~al.}(2015)Chen, Hu, Deng, and Xing]{nte3}
Chen,~J.; Hu,~L.; Deng,~J.; Xing,~X. Negative thermal expansion in functional materials: controllable thermal expansion by chemical modifications. \emph{Chem. Soc. Rev.} \textbf{2015}, \emph{44}, 3522--3567\relax
\mciteBstWouldAddEndPuncttrue
\mciteSetBstMidEndSepPunct{\mcitedefaultmidpunct}
{\mcitedefaultendpunct}{\mcitedefaultseppunct}\relax
\EndOfBibitem
\bibitem[Yang \latin{et~al.}(2026)Yang, Xu, Al-Maythalony, Huang, Xu, Hua, Zhang, Wang, Cui, Zhu, \latin{et~al.} others]{nteapp}
Yang,~H.; Xu,~P.; Al-Maythalony,~B.~A.; Huang,~Y.; Xu,~M.; Hua,~Y.; Zhang,~W.; Wang,~X.; Cui,~S.; Zhu,~H.; others Negative-thermal-expansion particles enable high-performance and ultradurable thermoelectric modules. \emph{Sci. Adv.} \textbf{2026}, \emph{12}, eaed3783\relax
\mciteBstWouldAddEndPuncttrue
\mciteSetBstMidEndSepPunct{\mcitedefaultmidpunct}
{\mcitedefaultendpunct}{\mcitedefaultseppunct}\relax
\EndOfBibitem
\bibitem[Zhu \latin{et~al.}(2025)Zhu, Yao, Liu, Chen, Gao, and Wang]{nteapp1}
Zhu,~Y.; Yao,~G.; Liu,~R.; Chen,~J.; Gao,~K.; Wang,~M. A novel negative thermal expansion metamaterial for supersonic aerothermoelastic flutter suppression. \emph{Aerosp. Sci. Technol.} \textbf{2025}, \emph{165}, 110493\relax
\mciteBstWouldAddEndPuncttrue
\mciteSetBstMidEndSepPunct{\mcitedefaultmidpunct}
{\mcitedefaultendpunct}{\mcitedefaultseppunct}\relax
\EndOfBibitem
\bibitem[Dubey \latin{et~al.}(2024)Dubey, Mirhakimi, and Elbestawi]{nteapp2}
Dubey,~D.; Mirhakimi,~A.~S.; Elbestawi,~M.~A. Negative thermal expansion metamaterials: a review of design, fabrication, and applications. \emph{J. Manuf. Mater. Process.} \textbf{2024}, \emph{8}, 40\relax
\mciteBstWouldAddEndPuncttrue
\mciteSetBstMidEndSepPunct{\mcitedefaultmidpunct}
{\mcitedefaultendpunct}{\mcitedefaultseppunct}\relax
\EndOfBibitem
\bibitem[Qian \latin{et~al.}(2026)Qian, Long, Liu, Fan, Liao, Gao, Lu, Gao, Kang, Song, \latin{et~al.} others]{nteapp3}
Qian,~J.; Long,~F.; Liu,~Y.; Fan,~L.; Liao,~M.; Gao,~Q.; Lu,~H.; Gao,~D.; Kang,~L.; Song,~Y.; others Wide Temperature Zero Thermal Expansion in Al Matrix Composites with High Thermal Conductivity. \emph{Adv. Sci.} \textbf{2026}, e75748\relax
\mciteBstWouldAddEndPuncttrue
\mciteSetBstMidEndSepPunct{\mcitedefaultmidpunct}
{\mcitedefaultendpunct}{\mcitedefaultseppunct}\relax
\EndOfBibitem
\bibitem[Makouei \latin{et~al.}(2021)Makouei, Nuri, and Nuri]{nteapp4}
Makouei,~S.; Nuri,~H.; Nuri,~M. M.~R. NTE material employment in temperature resistive high sensitive fiber optic strain sensor development. \emph{Optik} \textbf{2021}, \emph{231}, 166395\relax
\mciteBstWouldAddEndPuncttrue
\mciteSetBstMidEndSepPunct{\mcitedefaultmidpunct}
{\mcitedefaultendpunct}{\mcitedefaultseppunct}\relax
\EndOfBibitem
\bibitem[Wu \latin{et~al.}(2008)Wu, Kobayashi, Halder, Peterson, Chapman, Lock, Southon, and Kepert]{wu2008negative}
Wu,~Y.; Kobayashi,~A.; Halder,~G.~J.; Peterson,~V.~K.; Chapman,~K.~W.; Lock,~N.; Southon,~P.~D.; Kepert,~C.~J. Negative Thermal Expansion in the Metal--Organic Framework Material Cu3 (1, 3, 5-benzenetricarboxylate) 2. \emph{Angew. Chem., Int. Ed.} \textbf{2008}, \emph{47}, 8929--8932\relax
\mciteBstWouldAddEndPuncttrue
\mciteSetBstMidEndSepPunct{\mcitedefaultmidpunct}
{\mcitedefaultendpunct}{\mcitedefaultseppunct}\relax
\EndOfBibitem
\bibitem[Dubbeldam \latin{et~al.}(2007)Dubbeldam, Walton, Ellis, and Snurr]{isoreticular}
Dubbeldam,~D.; Walton,~K.~S.; Ellis,~D.~E.; Snurr,~R.~Q. Exceptional negative thermal expansion in isoreticular metal--organic frameworks. \emph{Angew. Chem.} \textbf{2007}, \emph{119}, 4580--4583\relax
\mciteBstWouldAddEndPuncttrue
\mciteSetBstMidEndSepPunct{\mcitedefaultmidpunct}
{\mcitedefaultendpunct}{\mcitedefaultseppunct}\relax
\EndOfBibitem
\bibitem[Zhou \latin{et~al.}(2008)Zhou, Wu, Yildirim, Simpson, and Walker]{zhou2008origin}
Zhou,~W.; Wu,~H.; Yildirim,~T.; Simpson,~J.~R.; Walker,~A.~H. Origin of the exceptional negative thermal expansion in metal-organic framework-5 Zn 4 O (1, 4-benzenedicarboxylate) 3. \emph{Phys. Rev. B:Condens. Matter Mater. Phys.} \textbf{2008}, \emph{78}, 054114\relax
\mciteBstWouldAddEndPuncttrue
\mciteSetBstMidEndSepPunct{\mcitedefaultmidpunct}
{\mcitedefaultendpunct}{\mcitedefaultseppunct}\relax
\EndOfBibitem
\bibitem[Zhou \latin{et~al.}(2012)Zhou, Long, and Yaghi]{mof}
Zhou,~H.-C.; Long,~J.~R.; Yaghi,~O.~M. Introduction to Metal--Organic Frameworks. \emph{Chem. Rev.} \textbf{2012}, \emph{112}, 673--674\relax
\mciteBstWouldAddEndPuncttrue
\mciteSetBstMidEndSepPunct{\mcitedefaultmidpunct}
{\mcitedefaultendpunct}{\mcitedefaultseppunct}\relax
\EndOfBibitem
\bibitem[Furukawa \latin{et~al.}(2013)Furukawa, Cordova, O’Keeffe, and Yaghi]{mof1}
Furukawa,~H.; Cordova,~K.~E.; O’Keeffe,~M.; Yaghi,~O.~M. The Chemistry and Applications of Metal-Organic Frameworks. \emph{Science} \textbf{2013}, \emph{341}, 1230444\relax
\mciteBstWouldAddEndPuncttrue
\mciteSetBstMidEndSepPunct{\mcitedefaultmidpunct}
{\mcitedefaultendpunct}{\mcitedefaultseppunct}\relax
\EndOfBibitem
\bibitem[Yaghi \latin{et~al.}(2003)Yaghi, O'Keeffe, Ockwig, Chae, Eddaoudi, and Kim]{mof2}
Yaghi,~O.~M.; O'Keeffe,~M.; Ockwig,~N.~W.; Chae,~H.~K.; Eddaoudi,~M.; Kim,~J. Reticular synthesis and the design of new materials. \emph{Nature} \textbf{2003}, \emph{423}, 705--714\relax
\mciteBstWouldAddEndPuncttrue
\mciteSetBstMidEndSepPunct{\mcitedefaultmidpunct}
{\mcitedefaultendpunct}{\mcitedefaultseppunct}\relax
\EndOfBibitem
\bibitem[Wieser \latin{et~al.}(2021)Wieser, Kamencek, Dürholt, Schmid, Bedoya-Martínez, and Zojer]{thermal}
Wieser,~S.; Kamencek,~T.; Dürholt,~J.~P.; Schmid,~R.; Bedoya-Martínez,~N.; Zojer,~E. Identifying the Bottleneck for Heat Transport in Metal–Organic Frameworks. \emph{Adv. Theory Simul.} \textbf{2021}, \emph{4}, 2000211\relax
\mciteBstWouldAddEndPuncttrue
\mciteSetBstMidEndSepPunct{\mcitedefaultmidpunct}
{\mcitedefaultendpunct}{\mcitedefaultseppunct}\relax
\EndOfBibitem
\bibitem[Hoffman \latin{et~al.}(2023)Hoffman, Senkovska, Abylgazina, Bon, Grzimek, Dominic, Russina, Kraft, Weidinger, Zeier, Van~Speybroeck, and Kaskel]{D3TA02214E}
Hoffman,~A. E.~J.; Senkovska,~I.; Abylgazina,~L.; Bon,~V.; Grzimek,~V.; Dominic,~A.~M.; Russina,~M.; Kraft,~M.~A.; Weidinger,~I.; Zeier,~W.~G.; Van~Speybroeck,~V.; Kaskel,~S. The role of phonons in switchable MOFs: a model material perspective. \emph{J. Mater. Chem. A} \textbf{2023}, \emph{11}, 15286--15300\relax
\mciteBstWouldAddEndPuncttrue
\mciteSetBstMidEndSepPunct{\mcitedefaultmidpunct}
{\mcitedefaultendpunct}{\mcitedefaultseppunct}\relax
\EndOfBibitem
\bibitem[Kuchta \latin{et~al.}(2019)Kuchta, Formalik, Rogacka, Neimark, and Firlej]{phonons}
Kuchta,~B.; Formalik,~F.; Rogacka,~J.; Neimark,~A.~V.; Firlej,~L. Phonons in deformable microporous crystalline solids. \emph{Z. Kristallogr. - Cryst. Mater.} \textbf{2019}, \emph{234}, 513--527\relax
\mciteBstWouldAddEndPuncttrue
\mciteSetBstMidEndSepPunct{\mcitedefaultmidpunct}
{\mcitedefaultendpunct}{\mcitedefaultseppunct}\relax
\EndOfBibitem
\bibitem[Ziman(2001)]{Ziman2001}
Ziman,~J.~M. \emph{Electrons and Phonons: The Theory of Transport Phenomena in Solids}; Oxford University Press, 2001; Vol. 540\relax
\mciteBstWouldAddEndPuncttrue
\mciteSetBstMidEndSepPunct{\mcitedefaultmidpunct}
{\mcitedefaultendpunct}{\mcitedefaultseppunct}\relax
\EndOfBibitem
\bibitem[Takenaka(2012)]{takenaka_negative_2012}
Takenaka,~K. Negative thermal expansion materials: technological key for control of thermal expansion. \emph{Sci. Technol. Adv. Mater.} \textbf{2012}, \emph{13}, 013001\relax
\mciteBstWouldAddEndPuncttrue
\mciteSetBstMidEndSepPunct{\mcitedefaultmidpunct}
{\mcitedefaultendpunct}{\mcitedefaultseppunct}\relax
\EndOfBibitem
\bibitem[Evans \latin{et~al.}(2019)Evans, D{\"u}rholt, Kaskel, and Schmid]{evans2019assessing}
Evans,~J.~D.; D{\"u}rholt,~J.~P.; Kaskel,~S.; Schmid,~R. Assessing negative thermal expansion in mesoporous metal--organic frameworks by molecular simulation. \emph{J. Mater. Chem. A} \textbf{2019}, \emph{7}, 24019--24026\relax
\mciteBstWouldAddEndPuncttrue
\mciteSetBstMidEndSepPunct{\mcitedefaultmidpunct}
{\mcitedefaultendpunct}{\mcitedefaultseppunct}\relax
\EndOfBibitem
\bibitem[Elena \latin{et~al.}(2025)Elena, Kamath, Jaffrelot~Inizan, Rosen, Zanca, and Persson]{kamath}
Elena,~A.~M.; Kamath,~P.~D.; Jaffrelot~Inizan,~T.; Rosen,~A.~S.; Zanca,~F.; Persson,~K.~A. Machine learned potential for high-throughput phonon calculations of metal--organic frameworks. \emph{npj Comput. Mater.} \textbf{2025}, \emph{11}, 125\relax
\mciteBstWouldAddEndPuncttrue
\mciteSetBstMidEndSepPunct{\mcitedefaultmidpunct}
{\mcitedefaultendpunct}{\mcitedefaultseppunct}\relax
\EndOfBibitem
\bibitem[Kamencek \latin{et~al.}(2022)Kamencek, Schrode, Resel, Ricco, and Zojer]{kamencek2022understanding}
Kamencek,~T.; Schrode,~B.; Resel,~R.; Ricco,~R.; Zojer,~E. Understanding the origin of the particularly small and anisotropic thermal expansion of MOF-74. \emph{Adv. Theory Simul.} \textbf{2022}, \emph{5}, 2200031\relax
\mciteBstWouldAddEndPuncttrue
\mciteSetBstMidEndSepPunct{\mcitedefaultmidpunct}
{\mcitedefaultendpunct}{\mcitedefaultseppunct}\relax
\EndOfBibitem
\bibitem[Ma \latin{et~al.}(2024)Ma, Liu, Chen, Li, Lin, Chen, Deng, Ohara, and Xing]{transition}
Ma,~R.; Liu,~Z.; Chen,~L.; Li,~Q.; Lin,~K.; Chen,~X.; Deng,~J.; Ohara,~K.; Xing,~X. Transition from isotropic positive to negative thermal expansion by local Zr6O8 node distortion in MOF-801. \emph{Microstructures} \textbf{2024}, \emph{4}, N--A\relax
\mciteBstWouldAddEndPuncttrue
\mciteSetBstMidEndSepPunct{\mcitedefaultmidpunct}
{\mcitedefaultendpunct}{\mcitedefaultseppunct}\relax
\EndOfBibitem
\bibitem[Burtch \latin{et~al.}(2019)Burtch, Baxter, Heinen, Bird, Schneemann, Dubbeldam, and Wilkinson]{burtch2019negative}
Burtch,~N.~C.; Baxter,~S.~J.; Heinen,~J.; Bird,~A.; Schneemann,~A.; Dubbeldam,~D.; Wilkinson,~A.~P. Negative thermal expansion design strategies in a diverse series of metal--organic frameworks. \emph{Adv. Funct. Mater.} \textbf{2019}, \emph{29}, 1904669\relax
\mciteBstWouldAddEndPuncttrue
\mciteSetBstMidEndSepPunct{\mcitedefaultmidpunct}
{\mcitedefaultendpunct}{\mcitedefaultseppunct}\relax
\EndOfBibitem
\bibitem[Baxter \latin{et~al.}(2019)Baxter, Schneemann, Ready, Wijeratne, Wilkinson, and Burtch]{baxter2019tuning}
Baxter,~S.~J.; Schneemann,~A.; Ready,~A.~D.; Wijeratne,~P.; Wilkinson,~A.~P.; Burtch,~N.~C. Tuning thermal expansion in metal--organic frameworks using a mixed linker solid solution approach. \emph{J. Am. Chem. Soc.} \textbf{2019}, \emph{141}, 12849--12854\relax
\mciteBstWouldAddEndPuncttrue
\mciteSetBstMidEndSepPunct{\mcitedefaultmidpunct}
{\mcitedefaultendpunct}{\mcitedefaultseppunct}\relax
\EndOfBibitem
\bibitem[Cliffe \latin{et~al.}(2015)Cliffe, Hill, Murray, Coudert, and Goodwin]{defect}
Cliffe,~M.~J.; Hill,~J.~A.; Murray,~C.~A.; Coudert,~F.-X.; Goodwin,~A.~L. Defect-dependent colossal negative thermal expansion in UiO-66 (Hf) metal--organic framework. \emph{Phys. Chem. Chem. Phys.} \textbf{2015}, \emph{17}, 11586--11592\relax
\mciteBstWouldAddEndPuncttrue
\mciteSetBstMidEndSepPunct{\mcitedefaultmidpunct}
{\mcitedefaultendpunct}{\mcitedefaultseppunct}\relax
\EndOfBibitem
\bibitem[Schneider \latin{et~al.}(2019)Schneider, Bodesheim, Ehrenreich, Crocell{\`a}, Mink, Fischer, Butler, and Kieslich]{retrofitting}
Schneider,~C.; Bodesheim,~D.; Ehrenreich,~M.~G.; Crocell{\`a},~V.; Mink,~J.; Fischer,~R.~A.; Butler,~K.~T.; Kieslich,~G. Tuning the negative thermal expansion behavior of the metal--organic framework Cu3BTC2 by retrofitting. \emph{J. Am. Chem. Soc.} \textbf{2019}, \emph{141}, 10504--10509\relax
\mciteBstWouldAddEndPuncttrue
\mciteSetBstMidEndSepPunct{\mcitedefaultmidpunct}
{\mcitedefaultendpunct}{\mcitedefaultseppunct}\relax
\EndOfBibitem
\bibitem[Vornholt \latin{et~al.}(2024)Vornholt, Chen, Hofmann, and Chapman]{vornholt2024node}
Vornholt,~S.~M.; Chen,~Z.; Hofmann,~J.; Chapman,~K.~W. Node distortions in UiO-66 inform negative thermal expansion mechanisms: kinetic effects, frustration, and lattice hysteresis. \emph{J. Am. Chem. Soc.} \textbf{2024}, \emph{146}, 16977--16981\relax
\mciteBstWouldAddEndPuncttrue
\mciteSetBstMidEndSepPunct{\mcitedefaultmidpunct}
{\mcitedefaultendpunct}{\mcitedefaultseppunct}\relax
\EndOfBibitem
\bibitem[Yue \latin{et~al.}(2024)Yue, Mohamed, and Jiang]{yue2024classifying}
Yue,~Y.; Mohamed,~S.~A.; Jiang,~J. Classifying and predicting the thermal expansion properties of metal--organic frameworks: a data-driven approach. \emph{J. Chem. Inf. Model.} \textbf{2024}, \emph{64}, 4966--4979\relax
\mciteBstWouldAddEndPuncttrue
\mciteSetBstMidEndSepPunct{\mcitedefaultmidpunct}
{\mcitedefaultendpunct}{\mcitedefaultseppunct}\relax
\EndOfBibitem
\bibitem[Dove \latin{et~al.}(2023)Dove, Wei, Phillips, Keen, and Refson]{rums}
Dove,~M.~T.; Wei,~Z.; Phillips,~A.~E.; Keen,~D.~A.; Refson,~K. Which phonons contribute most to negative thermal expansion in ScF3? \emph{APL Mater.} \textbf{2023}, \emph{11}\relax
\mciteBstWouldAddEndPuncttrue
\mciteSetBstMidEndSepPunct{\mcitedefaultmidpunct}
{\mcitedefaultendpunct}{\mcitedefaultseppunct}\relax
\EndOfBibitem
\bibitem[Rappe \latin{et~al.}(1992)Rappe, Casewit, Colwell, Goddard, and Skiff]{uff}
Rappe,~A.~K.; Casewit,~C.~J.; Colwell,~K.~S.; Goddard,~W. A.~I.; Skiff,~W.~M. UFF, a full periodic table force field for molecular mechanics and molecular dynamics simulations. \emph{J. Am. Chem. Soc.} \textbf{1992}, \emph{114}, 10024--10035\relax
\mciteBstWouldAddEndPuncttrue
\mciteSetBstMidEndSepPunct{\mcitedefaultmidpunct}
{\mcitedefaultendpunct}{\mcitedefaultseppunct}\relax
\EndOfBibitem
\bibitem[Bureekaew \latin{et~al.}(2013)Bureekaew, Amirjalayer, Tafipolsky, Spickermann, Roy, and Schmid]{mofff}
Bureekaew,~S.; Amirjalayer,~S.; Tafipolsky,~M.; Spickermann,~C.; Roy,~T.~K.; Schmid,~R. MOF-FF – A flexible first-principles derived force field for metal-organic frameworks. \emph{Phys. Status Solidi B} \textbf{2013}, \emph{250}, 1128--1141\relax
\mciteBstWouldAddEndPuncttrue
\mciteSetBstMidEndSepPunct{\mcitedefaultmidpunct}
{\mcitedefaultendpunct}{\mcitedefaultseppunct}\relax
\EndOfBibitem
\bibitem[Wieser and Zojer(2024)Wieser, and Zojer]{Wieser2024}
Wieser,~S.; Zojer,~E. Machine learned force-fields for an Ab-initio quality description of metal--organic frameworks. \emph{npj Comput. Mater.} \textbf{2024}, \emph{10}, 18\relax
\mciteBstWouldAddEndPuncttrue
\mciteSetBstMidEndSepPunct{\mcitedefaultmidpunct}
{\mcitedefaultendpunct}{\mcitedefaultseppunct}\relax
\EndOfBibitem
\bibitem[Bristow \latin{et~al.}(2016)Bristow, Skelton, Svane, Walsh, and Gale]{vmof}
Bristow,~J.~K.; Skelton,~J.~M.; Svane,~K.~L.; Walsh,~A.; Gale,~J.~D. A general forcefield for accurate phonon properties of metal–organic frameworks. \emph{Phys. Chem. Chem. Phys.} \textbf{2016}, \emph{18}, 29316--29329\relax
\mciteBstWouldAddEndPuncttrue
\mciteSetBstMidEndSepPunct{\mcitedefaultmidpunct}
{\mcitedefaultendpunct}{\mcitedefaultseppunct}\relax
\EndOfBibitem
\bibitem[Moosavi \latin{et~al.}(2022)Moosavi, Novotny, Ongari, Moubarak, Asgari, Kadioglu, Charalambous, Ortega-Guerrero, Farmahini, Sarkisov, \latin{et~al.} others]{Moosavi}
Moosavi,~S.~M.; Novotny,~B.~{\'A}.; Ongari,~D.; Moubarak,~E.; Asgari,~M.; Kadioglu,~{\"O}.; Charalambous,~C.; Ortega-Guerrero,~A.; Farmahini,~A.~H.; Sarkisov,~L.; others A data-science approach to predict the heat capacity of nanoporous materials. \emph{Nat. Mater.} \textbf{2022}, \emph{21}, 1419--1425\relax
\mciteBstWouldAddEndPuncttrue
\mciteSetBstMidEndSepPunct{\mcitedefaultmidpunct}
{\mcitedefaultendpunct}{\mcitedefaultseppunct}\relax
\EndOfBibitem
\bibitem[Yue \latin{et~al.}(2025)Yue, Mohamed, Loh, and Jiang]{yue}
Yue,~Y.; Mohamed,~S.~A.; Loh,~N.~D.; Jiang,~J. Toward a Generalizable Machine-Learned Potential for Metal--Organic Frameworks. \emph{ACS Nano} \textbf{2025}, \emph{19}, 933--949\relax
\mciteBstWouldAddEndPuncttrue
\mciteSetBstMidEndSepPunct{\mcitedefaultmidpunct}
{\mcitedefaultendpunct}{\mcitedefaultseppunct}\relax
\EndOfBibitem
\bibitem[Kamath and Persson(2026)Kamath, and Persson]{spurious}
Kamath,~P.~D.; Persson,~K.~A. The impact of spurious imaginary phonon modes on thermal properties of Metal-organic Frameworks. \emph{npj Comput. Mat.} \textbf{2026}, \emph{12}, 378\relax
\mciteBstWouldAddEndPuncttrue
\mciteSetBstMidEndSepPunct{\mcitedefaultmidpunct}
{\mcitedefaultendpunct}{\mcitedefaultseppunct}\relax
\EndOfBibitem
\bibitem[Loew \latin{et~al.}(2025)Loew, Sun, Wang, Botti, and Marques]{loew2025universal}
Loew,~A.; Sun,~D.; Wang,~H.-C.; Botti,~S.; Marques,~M.~A. Universal machine learning interatomic potentials are ready for phonons. \emph{npj Comput. Mater.} \textbf{2025}, \emph{11}, 178\relax
\mciteBstWouldAddEndPuncttrue
\mciteSetBstMidEndSepPunct{\mcitedefaultmidpunct}
{\mcitedefaultendpunct}{\mcitedefaultseppunct}\relax
\EndOfBibitem
\bibitem[Kra{\ss} \latin{et~al.}(2025)Kra{\ss}, Huang, and Moosavi]{mofsim}
Kra{\ss},~H.; Huang,~J.; Moosavi,~S.~M. MOFSimBench: evaluating universal machine learning interatomic potentials in metal-organic framework molecular modeling. \emph{npj Comput. Mater.} \textbf{2025}, \emph{12}, 4\relax
\mciteBstWouldAddEndPuncttrue
\mciteSetBstMidEndSepPunct{\mcitedefaultmidpunct}
{\mcitedefaultendpunct}{\mcitedefaultseppunct}\relax
\EndOfBibitem
\bibitem[Batatia \latin{et~al.}(2025)Batatia, Benner, Chiang, Elena, Kovács, Riebesell, Advincula, Asta, Avaylon, Baldwin, Berger, Bernstein, Bhowmik, Bigi, Blau, Cărare, Ceriotti, Chong, Darby, De, Della~Pia, Deringer, Elijošius, El-Machachi, Fako, Falcioni, Ferrari, Gardner, Gawkowski, Genreith-Schriever, George, Goodall, Grandel, Grey, Grigorev, Han, Handley, Heenen, Hermansson, Ho, Hofmann, Holm, Jaafar, Jakob, Jung, Kapil, Kaplan, Karimitari, Kermode, Kourtis, Kroupa, Kullgren, Kuner, Kuryla, Liepuoniute, Lin, Margraf, Magdău, Michaelides, Moore, Naik, Niblett, Norwood, O’Neill, Ortner, Persson, Reuter, Rosen, Rosset, Schaaf, Schran, Shi, Sivonxay, Stenczel, Sutton, Svahn, Swinburne, Tilly, van~der Oord, Vargas, Varga-Umbrich, Vegge, Vondrák, Wang, Witt, Wolf, Zills, and Csányi]{mace}
Batatia,~I. \latin{et~al.}  A foundation model for atomistic materials chemistry. \emph{J. Chem. Phys} \textbf{2025}, \emph{163}, 184110\relax
\mciteBstWouldAddEndPuncttrue
\mciteSetBstMidEndSepPunct{\mcitedefaultmidpunct}
{\mcitedefaultendpunct}{\mcitedefaultseppunct}\relax
\EndOfBibitem
\bibitem[Rosen \latin{et~al.}(2021)Rosen, Iyer, Ray, Yao, Aspuru-Guzik, Gagliardi, Notestein, and Snurr]{qmof1}
Rosen,~A.~S.; Iyer,~S.~M.; Ray,~D.; Yao,~Z.; Aspuru-Guzik,~A.; Gagliardi,~L.; Notestein,~J.~M.; Snurr,~R.~Q. Machine learning the quantum-chemical properties of metal--organic frameworks for accelerated materials discovery. \emph{Matter} \textbf{2021}, \emph{4}, 1578--1597\relax
\mciteBstWouldAddEndPuncttrue
\mciteSetBstMidEndSepPunct{\mcitedefaultmidpunct}
{\mcitedefaultendpunct}{\mcitedefaultseppunct}\relax
\EndOfBibitem
\bibitem[Rosen \latin{et~al.}(2022)Rosen, Fung, Huck, O'Donnell, Horton, Truhlar, Persson, Notestein, and Snurr]{qmof2}
Rosen,~A.~S.; Fung,~V.; Huck,~P.; O'Donnell,~C.~T.; Horton,~M.~K.; Truhlar,~D.~G.; Persson,~K.~A.; Notestein,~J.~M.; Snurr,~R.~Q. High-throughput predictions of metal--organic framework electronic properties: theoretical challenges, graph neural networks, and data exploration. \emph{npj Comput. Mater.} \textbf{2022}, \emph{8}, 112\relax
\mciteBstWouldAddEndPuncttrue
\mciteSetBstMidEndSepPunct{\mcitedefaultmidpunct}
{\mcitedefaultendpunct}{\mcitedefaultseppunct}\relax
\EndOfBibitem
\bibitem[White \latin{et~al.}(2025)White, Gibaldi, Burner, Mayo, and Woo]{white_high_2025}
White,~A.~J.; Gibaldi,~M.; Burner,~J.; Mayo,~R.~A.; Woo,~T.~K. High {Structural} {Error} {Rates} in “{Computation}-{Ready}” {MOF} {Databases} {Discovered} by {Checking} {Metal} {Oxidation} {States}. \emph{J. Am. Chem. Soc.} \textbf{2025}, \emph{147}, 17579--17583\relax
\mciteBstWouldAddEndPuncttrue
\mciteSetBstMidEndSepPunct{\mcitedefaultmidpunct}
{\mcitedefaultendpunct}{\mcitedefaultseppunct}\relax
\EndOfBibitem
\bibitem[Chung \latin{et~al.}(2019)Chung, Haldoupis, Bucior, Haranczyk, Lee, Zhang, Vogiatzis, Milisavljevic, Ling, Camp, Slater, Siepmann, Sholl, and Snurr]{chung_advances_2019}
Chung,~Y.~G.; Haldoupis,~E.; Bucior,~B.~J.; Haranczyk,~M.; Lee,~S.; Zhang,~H.; Vogiatzis,~K.~D.; Milisavljevic,~M.; Ling,~S.; Camp,~J.~S.; Slater,~B.; Siepmann,~J.~I.; Sholl,~D.~S.; Snurr,~R.~Q. Advances, {Updates}, and {Analytics} for the {Computation}-{Ready}, {Experimental} {Metal}–{Organic} {Framework} {Database}: {CoRE} {MOF} 2019. \emph{J. Chem. Eng. Data.} \textbf{2019}, \emph{64}, 5985--5998\relax
\mciteBstWouldAddEndPuncttrue
\mciteSetBstMidEndSepPunct{\mcitedefaultmidpunct}
{\mcitedefaultendpunct}{\mcitedefaultseppunct}\relax
\EndOfBibitem
\bibitem[Togo \latin{et~al.}(2023)Togo, Chaput, Tadano, and Tanaka]{phonopy-phono3py-JPCM}
Togo,~A.; Chaput,~L.; Tadano,~T.; Tanaka,~I. Implementation strategies in phonopy and phono3py. \emph{J. Phys. Condens. Matter} \textbf{2023}, \emph{35}, 353001\relax
\mciteBstWouldAddEndPuncttrue
\mciteSetBstMidEndSepPunct{\mcitedefaultmidpunct}
{\mcitedefaultendpunct}{\mcitedefaultseppunct}\relax
\EndOfBibitem
\bibitem[Togo(2023)]{phonopy-phono3py-JPSJ}
Togo,~A. First-principles Phonon Calculations with Phonopy and Phono3py. \emph{J. Phys. Soc. Jpn.} \textbf{2023}, \emph{92}, 012001\relax
\mciteBstWouldAddEndPuncttrue
\mciteSetBstMidEndSepPunct{\mcitedefaultmidpunct}
{\mcitedefaultendpunct}{\mcitedefaultseppunct}\relax
\EndOfBibitem
\bibitem[Dymkowski \latin{et~al.}(2018)Dymkowski, Parker, Fern{\'a}ndez-Alonso, and Mukhopadhyay]{Dymkowski2018AbINS}
Dymkowski,~K.; Parker,~S.~F.; Fern{\'a}ndez-Alonso,~F.; Mukhopadhyay,~S. AbINS: The modern software for {INS} interpretation. \emph{Phys. B} \textbf{2018}, \emph{551}, 443--448\relax
\mciteBstWouldAddEndPuncttrue
\mciteSetBstMidEndSepPunct{\mcitedefaultmidpunct}
{\mcitedefaultendpunct}{\mcitedefaultseppunct}\relax
\EndOfBibitem
\bibitem[Arnold \latin{et~al.}(2014)Arnold, Bilheux, Borreguero, Buts, Campbell, Chapon, Doucet, Draper, Leal, Gigg, Lynch, Markvardsen, Mikkelson, Mikkelson, Miller, Palmen, Parker, Passos, Perring, Peterson, Ren, Reuter, Savici, Taylor, Taylor, Tolchenov, Zhou, and Zikovsky]{arnoldMantidDataAnalysis2014}
Arnold,~O. \latin{et~al.}  Mantid {Data} analysis and visualization package for neutron scattering and $\mu${SR} experiments. \emph{Nucl. Instrum. Methods Phys. Res., Sect. A} \textbf{2014}, \emph{764}, 156--166\relax
\mciteBstWouldAddEndPuncttrue
\mciteSetBstMidEndSepPunct{\mcitedefaultmidpunct}
{\mcitedefaultendpunct}{\mcitedefaultseppunct}\relax
\EndOfBibitem
\bibitem[Pinna \latin{et~al.}(2018)Pinna, Rudi{\'c}, Parker, Armstrong, Zanetti, {\v S}koro, Waller, Zacek, Smith, Capstick, McPhail, Pooley, Howells, Gorini, and Fernandez-Alonso]{Pinna2018NeutronGuideTOSCA}
Pinna,~R.~S.; Rudi{\'c},~S.; Parker,~S.~F.; Armstrong,~J.; Zanetti,~M.; {\v S}koro,~G.; Waller,~S.~P.; Zacek,~D.; Smith,~C.~A.; Capstick,~M.~J.; McPhail,~D.~J.; Pooley,~D.~E.; Howells,~G.~D.; Gorini,~G.; Fernandez-Alonso,~F. The neutron guide upgrade of the TOSCA spectrometer. \emph{Nucl. Instrum. Methods Phys. Res., Sect. A} \textbf{2018}, \emph{896}, 68--74\relax
\mciteBstWouldAddEndPuncttrue
\mciteSetBstMidEndSepPunct{\mcitedefaultmidpunct}
{\mcitedefaultendpunct}{\mcitedefaultseppunct}\relax
\EndOfBibitem
\bibitem[Kresse and Furthm\"uller(1996)Kresse, and Furthm\"uller]{vasp}
Kresse,~G.; Furthm\"uller,~J. Efficient iterative schemes for ab initio total-energy calculations using a plane-wave basis set. \emph{Phys. Rev. B} \textbf{1996}, \emph{54}, 11169--11186\relax
\mciteBstWouldAddEndPuncttrue
\mciteSetBstMidEndSepPunct{\mcitedefaultmidpunct}
{\mcitedefaultendpunct}{\mcitedefaultseppunct}\relax
\EndOfBibitem
\bibitem[Kresse and Joubert(1999)Kresse, and Joubert]{vasp1}
Kresse,~G.; Joubert,~D. From ultrasoft pseudopotentials to the projector augmented-wave method. \emph{Phys. Rev. B} \textbf{1999}, \emph{59}, 1758--1775\relax
\mciteBstWouldAddEndPuncttrue
\mciteSetBstMidEndSepPunct{\mcitedefaultmidpunct}
{\mcitedefaultendpunct}{\mcitedefaultseppunct}\relax
\EndOfBibitem
\bibitem[Kresse and Furthmüller(1996)Kresse, and Furthmüller]{vasp2}
Kresse,~G.; Furthmüller,~J. Efficiency of ab-initio total energy calculations for metals and semiconductors using a plane-wave basis set. \emph{Comput. Mater. Sci.} \textbf{1996}, \emph{6}, 15--50\relax
\mciteBstWouldAddEndPuncttrue
\mciteSetBstMidEndSepPunct{\mcitedefaultmidpunct}
{\mcitedefaultendpunct}{\mcitedefaultseppunct}\relax
\EndOfBibitem
\bibitem[Kresse and Hafner(1993)Kresse, and Hafner]{vasp3}
Kresse,~G.; Hafner,~J. Ab initio molecular dynamics for liquid metals. \emph{Phys. Rev. B} \textbf{1993}, \emph{47}, 558--561\relax
\mciteBstWouldAddEndPuncttrue
\mciteSetBstMidEndSepPunct{\mcitedefaultmidpunct}
{\mcitedefaultendpunct}{\mcitedefaultseppunct}\relax
\EndOfBibitem
\bibitem[Perdew \latin{et~al.}(1996)Perdew, Burke, and Ernzerhof]{pbegga}
Perdew,~J.~P.; Burke,~K.; Ernzerhof,~M. Generalized Gradient Approximation Made Simple. \emph{Phys. Rev. Lett.} \textbf{1996}, \emph{77}, 3865--3868\relax
\mciteBstWouldAddEndPuncttrue
\mciteSetBstMidEndSepPunct{\mcitedefaultmidpunct}
{\mcitedefaultendpunct}{\mcitedefaultseppunct}\relax
\EndOfBibitem
\bibitem[Grimme \latin{et~al.}(2011)Grimme, Ehrlich, and Goerigk]{bj}
Grimme,~S.; Ehrlich,~S.; Goerigk,~L. Effect of the damping function in dispersion corrected density functional theory. \emph{J. Comput. Chem.} \textbf{2011}, \emph{32}, 1456--1465\relax
\mciteBstWouldAddEndPuncttrue
\mciteSetBstMidEndSepPunct{\mcitedefaultmidpunct}
{\mcitedefaultendpunct}{\mcitedefaultseppunct}\relax
\EndOfBibitem
\bibitem[Grimme \latin{et~al.}(2010)Grimme, Antony, Ehrlich, and Krieg]{d3}
Grimme,~S.; Antony,~J.; Ehrlich,~S.; Krieg,~H. {A consistent and accurate ab initio parametrization of density functional dispersion correction (DFT-D) for the 94 elements H-Pu}. \emph{J. Chem. Phys.} \textbf{2010}, \emph{132}, 154104\relax
\mciteBstWouldAddEndPuncttrue
\mciteSetBstMidEndSepPunct{\mcitedefaultmidpunct}
{\mcitedefaultendpunct}{\mcitedefaultseppunct}\relax
\EndOfBibitem
\bibitem[Gao \latin{et~al.}(2024)Gao, Feng, Ren, and Li]{GAO2024124597}
Gao,~L.-L.; Feng,~J.-Y.; Ren,~H.-M.; Li,~G. High intrinsic proton conduction in cerium(IV) metal-organic framework constructed by bromine-grafted terephthalic acid. \emph{J. Solid State Chem.} \textbf{2024}, \emph{333}, 124597\relax
\mciteBstWouldAddEndPuncttrue
\mciteSetBstMidEndSepPunct{\mcitedefaultmidpunct}
{\mcitedefaultendpunct}{\mcitedefaultseppunct}\relax
\EndOfBibitem
\bibitem[Toby and Von~Dreele(2013)Toby, and Von~Dreele]{GSASII}
Toby,~B.~H.; Von~Dreele,~R.~B. GSAS-II: the genesis of a modern open-source all purpose crystallography software package. \emph{J. Appl. Crystallogr.} \textbf{2013}, \emph{46}, 544--549\relax
\mciteBstWouldAddEndPuncttrue
\mciteSetBstMidEndSepPunct{\mcitedefaultmidpunct}
{\mcitedefaultendpunct}{\mcitedefaultseppunct}\relax
\EndOfBibitem
\bibitem[Son \latin{et~al.}(2020)Son, Atilgan, Idrees, Islamoglu, and Farha]{son2020solvent}
Son,~F.~A.; Atilgan,~A.; Idrees,~K.~B.; Islamoglu,~T.; Farha,~O.~K. Solvent-assisted linker exchange enabled preparation of cerium-based metal--organic frameworks constructed from redox active linkers. \emph{Inorganic Chemistry Frontiers} \textbf{2020}, \emph{7}, 984--990\relax
\mciteBstWouldAddEndPuncttrue
\mciteSetBstMidEndSepPunct{\mcitedefaultmidpunct}
{\mcitedefaultendpunct}{\mcitedefaultseppunct}\relax
\EndOfBibitem
\bibitem[Gonz{\'a}lez \latin{et~al.}(2024)Gonz{\'a}lez, Pazo-Carballo, Cam{\'u}, Hidalgo-Rosa, Zarate, Escalona, and Schott]{gonzalez2024adsorption}
Gonz{\'a}lez,~D.; Pazo-Carballo,~C.; Cam{\'u},~E.; Hidalgo-Rosa,~Y.; Zarate,~X.; Escalona,~N.; Schott,~E. Adsorption properties of M-UiO-66 (M= Zr (iv); Hf (iv) or Ce (iv)) with BDC or PDC linker. \emph{Dalton Trans.} \textbf{2024}, \emph{53}, 10486--10498\relax
\mciteBstWouldAddEndPuncttrue
\mciteSetBstMidEndSepPunct{\mcitedefaultmidpunct}
{\mcitedefaultendpunct}{\mcitedefaultseppunct}\relax
\EndOfBibitem
\bibitem[Snurr(2019)]{SNURR201926}
Snurr,~R.~Q. It’s an Interesting MOF, but Is It Stable? \emph{Matter} \textbf{2019}, \emph{1}, 26--27\relax
\mciteBstWouldAddEndPuncttrue
\mciteSetBstMidEndSepPunct{\mcitedefaultmidpunct}
{\mcitedefaultendpunct}{\mcitedefaultseppunct}\relax
\EndOfBibitem
\bibitem[Bahadur \latin{et~al.}(2025)Bahadur, Sharma, Kancharlapalli, Mor, Armstrong, and Sen]{bahadur2025competitive}
Bahadur,~J.; Sharma,~S.~K.; Kancharlapalli,~S.; Mor,~J.; Armstrong,~J.; Sen,~D. Competitive Host--Guest and Guest--Guest Interaction Dependent Flexibility of Crystal-Size-Engineered ZIF-8: An Inelastic Neutron Scattering Study. \emph{J. Phys. Chem. Lett.} \textbf{2025}, \emph{16}, 5496--5505\relax
\mciteBstWouldAddEndPuncttrue
\mciteSetBstMidEndSepPunct{\mcitedefaultmidpunct}
{\mcitedefaultendpunct}{\mcitedefaultseppunct}\relax
\EndOfBibitem
\bibitem[Casco \latin{et~al.}(2016)Casco, Rey, Jord{\'a}, Rudi{\'c}, Fauth, Martinez-Escandell, Rodriguez-Reinoso, Ramos-Fern{\'a}ndez, and Silvestre-Albero]{casco2016paving}
Casco,~M.~E.; Rey,~F.; Jord{\'a},~J.~L.; Rudi{\'c},~S.; Fauth,~F.; Martinez-Escandell,~M.; Rodriguez-Reinoso,~F.; Ramos-Fern{\'a}ndez,~E.~V.; Silvestre-Albero,~J. Paving the way for methane hydrate formation on metal--organic frameworks (MOFs). \emph{Chem. Sci.} \textbf{2016}, \emph{7}, 3658--3666\relax
\mciteBstWouldAddEndPuncttrue
\mciteSetBstMidEndSepPunct{\mcitedefaultmidpunct}
{\mcitedefaultendpunct}{\mcitedefaultseppunct}\relax
\EndOfBibitem
\bibitem[Callear \latin{et~al.}(2013)Callear, Ramirez-Cuesta, David, Millange, and Walton]{callear2013high}
Callear,~S.~K.; Ramirez-Cuesta,~A.~J.; David,~W.~I.; Millange,~F.; Walton,~R.~I. High-resolution inelastic neutron scattering and neutron powder diffraction study of the adsorption of dihydrogen by the Cu (II) metal--organic framework material HKUST-1. \emph{Chem. Phys.} \textbf{2013}, \emph{427}, 9--17\relax
\mciteBstWouldAddEndPuncttrue
\mciteSetBstMidEndSepPunct{\mcitedefaultmidpunct}
{\mcitedefaultendpunct}{\mcitedefaultseppunct}\relax
\EndOfBibitem
\bibitem[Chen \latin{et~al.}(2023)Chen, Lu, Schroder, and Yang]{chen2023analysis}
Chen,~Y.; Lu,~W.; Schroder,~M.; Yang,~S. Analysis and refinement of host--guest interactions in metal--organic frameworks. \emph{Acc. Chem. Res.} \textbf{2023}, \emph{56}, 2569--2581\relax
\mciteBstWouldAddEndPuncttrue
\mciteSetBstMidEndSepPunct{\mcitedefaultmidpunct}
{\mcitedefaultendpunct}{\mcitedefaultseppunct}\relax
\EndOfBibitem
\bibitem[Inizan \latin{et~al.}(2026)Inizan, Kamath, Elena, and Persson]{umof}
Inizan,~T.~J.; Kamath,~P.~D.; Elena,~A.~M.; Persson,~K.~A. uMOF: A Universal Database, Benchmark, and Machine Learning Interatomic Potentials for Metal-Organic Frameworks. \emph{arXiv preprint arXiv:2608.28100} \textbf{2026}, \relax
\mciteBstWouldAddEndPunctfalse
\mciteSetBstMidEndSepPunct{\mcitedefaultmidpunct}
{}{\mcitedefaultseppunct}\relax
\EndOfBibitem
\bibitem[Granhed \latin{et~al.}(2019)Granhed, Lindman, Ekl{\"o}f-{\"O}sterberg, Karlsson, Parker, and Wahnstr{\"o}m]{under}
Granhed,~E.~J.; Lindman,~A.; Ekl{\"o}f-{\"O}sterberg,~C.; Karlsson,~M.; Parker,~S.~F.; Wahnstr{\"o}m,~G. Band vs. polaron: vibrational motion and chemical expansion of hydride ions as signatures for the electronic character in oxyhydride barium titanate. \emph{J. Mater. Chem. A} \textbf{2019}, \emph{7}, 16211--16221\relax
\mciteBstWouldAddEndPuncttrue
\mciteSetBstMidEndSepPunct{\mcitedefaultmidpunct}
{\mcitedefaultendpunct}{\mcitedefaultseppunct}\relax
\EndOfBibitem
\bibitem[Franchini \latin{et~al.}(2010)Franchini, Sanna, Marsman, and Kresse]{under1}
Franchini,~C.; Sanna,~A.; Marsman,~M.; Kresse,~G. Structural, vibrational, and quasiparticle properties of the Peierls semiconductor BaBiO 3: A hybrid functional and self-consistent GW+ vertex-corrections study. \emph{Phys. Rev. B:Condens. Matter Mater. Phys.} \textbf{2010}, \emph{81}, 085213\relax
\mciteBstWouldAddEndPuncttrue
\mciteSetBstMidEndSepPunct{\mcitedefaultmidpunct}
{\mcitedefaultendpunct}{\mcitedefaultseppunct}\relax
\EndOfBibitem
\bibitem[Cheng \latin{et~al.}(2023)Cheng, Stone, and Ramirez-Cuesta]{cheng2023database}
Cheng,~Y.; Stone,~M.~B.; Ramirez-Cuesta,~A.~J. A database of synthetic inelastic neutron scattering spectra from molecules and crystals. \emph{Sci. Data} \textbf{2023}, \emph{10}, 54\relax
\mciteBstWouldAddEndPuncttrue
\mciteSetBstMidEndSepPunct{\mcitedefaultmidpunct}
{\mcitedefaultendpunct}{\mcitedefaultseppunct}\relax
\EndOfBibitem
\bibitem[Henke \latin{et~al.}(2013)Henke, Schneemann, and Fischer]{massive}
Henke,~S.; Schneemann,~A.; Fischer,~R.~A. Massive anisotropic thermal expansion and thermo-responsive breathing in metal--organic frameworks modulated by linker functionalization. \emph{Adv. Funct. Mater.} \textbf{2013}, \emph{23}, 5990--5996\relax
\mciteBstWouldAddEndPuncttrue
\mciteSetBstMidEndSepPunct{\mcitedefaultmidpunct}
{\mcitedefaultendpunct}{\mcitedefaultseppunct}\relax
\EndOfBibitem
\bibitem[Liu \latin{et~al.}(2018)Liu, Li, Zhu, Lin, Deng, Chen, and Xing]{ortho}
Liu,~Z.; Li,~Q.; Zhu,~H.; Lin,~K.; Deng,~J.; Chen,~J.; Xing,~X. 3D negative thermal expansion in orthorhombic MIL-68 (In). \emph{Chem. Commun.} \textbf{2018}, \emph{54}, 5712--5715\relax
\mciteBstWouldAddEndPuncttrue
\mciteSetBstMidEndSepPunct{\mcitedefaultmidpunct}
{\mcitedefaultendpunct}{\mcitedefaultseppunct}\relax
\EndOfBibitem
\bibitem[Linnera \latin{et~al.}(2019)Linnera, Erba, and Karttunen]{linnera2019negative}
Linnera,~J.; Erba,~A.; Karttunen,~A.~J. Negative thermal expansion of Cu2O studied by quasi-harmonic approximation and cubic force-constant method. \emph{J. Chem. Phys.} \textbf{2019}, \emph{151}\relax
\mciteBstWouldAddEndPuncttrue
\mciteSetBstMidEndSepPunct{\mcitedefaultmidpunct}
{\mcitedefaultendpunct}{\mcitedefaultseppunct}\relax
\EndOfBibitem
\bibitem[Lock \latin{et~al.}(2013)Lock, Christensen, Wu, Peterson, Thomsen, Piltz, Ramirez-Cuesta, McIntyre, Nor{\'e}n, Kutteh, \latin{et~al.} others]{insmof-5}
Lock,~N.; Christensen,~M.; Wu,~Y.; Peterson,~V.~K.; Thomsen,~M.~K.; Piltz,~R.~O.; Ramirez-Cuesta,~A.~J.; McIntyre,~G.~J.; Nor{\'e}n,~K.; Kutteh,~R.; others Scrutinizing negative thermal expansion in MOF-5 by scattering techniques and ab initio calculations. \emph{Dalton Trans.} \textbf{2013}, \emph{42}, 1996--2007\relax
\mciteBstWouldAddEndPuncttrue
\mciteSetBstMidEndSepPunct{\mcitedefaultmidpunct}
{\mcitedefaultendpunct}{\mcitedefaultseppunct}\relax
\EndOfBibitem
\bibitem[Liu \latin{et~al.}(2023)Liu, Chen, Huang, Liu, Wang, Lan, Wei, and Wang]{imaging}
Liu,~B.; Chen,~X.; Huang,~N.; Liu,~S.; Wang,~Y.; Lan,~X.; Wei,~F.; Wang,~T. Imaging the dynamic influence of functional groups on metal-organic frameworks. \emph{Nat. Commun.} \textbf{2023}, \emph{14}, 4835\relax
\mciteBstWouldAddEndPuncttrue
\mciteSetBstMidEndSepPunct{\mcitedefaultmidpunct}
{\mcitedefaultendpunct}{\mcitedefaultseppunct}\relax
\EndOfBibitem
\bibitem[Taddei \latin{et~al.}(2017)Taddei, Tiana, Casati, Van~Bokhoven, Smit, and Ranocchiari]{taddei2017mixed}
Taddei,~M.; Tiana,~D.; Casati,~N.; Van~Bokhoven,~J.~A.; Smit,~B.; Ranocchiari,~M. Mixed-linker UiO-66: structure--property relationships revealed by a combination of high-resolution powder X-ray diffraction and density functional theory calculations. \emph{Phys. Chem. Chem. Phys.} \textbf{2017}, \emph{19}, 1551--1559\relax
\mciteBstWouldAddEndPuncttrue
\mciteSetBstMidEndSepPunct{\mcitedefaultmidpunct}
{\mcitedefaultendpunct}{\mcitedefaultseppunct}\relax
\EndOfBibitem
\bibitem[Goodenough \latin{et~al.}(2020)Goodenough, Devulapalli, Xu, Boyanich, Luo, De~Souza, Richard, Rosi, and Borguet]{goodenough2020interplay}
Goodenough,~I.; Devulapalli,~V. S.~D.; Xu,~W.; Boyanich,~M.~C.; Luo,~T.-Y.; De~Souza,~M.; Richard,~M.; Rosi,~N.~L.; Borguet,~E. Interplay between intrinsic thermal stability and expansion properties of functionalized UiO-67 metal--organic frameworks. \emph{Chem. Mater.} \textbf{2020}, \emph{33}, 910--920\relax
\mciteBstWouldAddEndPuncttrue
\mciteSetBstMidEndSepPunct{\mcitedefaultmidpunct}
{\mcitedefaultendpunct}{\mcitedefaultseppunct}\relax
\EndOfBibitem
\bibitem[Tavani \latin{et~al.}(2025)Tavani, Tofoni, Pietropaoli, Stoian, van Beek, Marshall, Pettiti, Latini, and D’Angelo]{MOF1f}
Tavani,~F.; Tofoni,~A.; Pietropaoli,~E.; Stoian,~D.~C.; van Beek,~W.; Marshall,~K.; Pettiti,~I.; Latini,~A.; D’Angelo,~P. Decoding the Water Harvesting Mechanism of MIL-100(Fe) Across Short- and Long-Range Length Scales. \emph{J. Am. Chem. Soc.} \textbf{2025}, \emph{147}, 40507--40518\relax
\mciteBstWouldAddEndPuncttrue
\mciteSetBstMidEndSepPunct{\mcitedefaultmidpunct}
{\mcitedefaultendpunct}{\mcitedefaultseppunct}\relax
\EndOfBibitem
\bibitem[Tavani \latin{et~al.}(2025)Tavani, Tofoni, Vandone, Busato, Braglia, Torelli, Stanzione, Armstrong, Morris, Colombo, and D'Angelo]{MOF2f}
Tavani,~F.; Tofoni,~A.; Vandone,~M.; Busato,~M.; Braglia,~L.; Torelli,~P.; Stanzione,~M.~G.; Armstrong,~A.~R.; Morris,~R.~E.; Colombo,~V.; D'Angelo,~P. A combined soft X-ray and theoretical investigation discloses the water harvesting behaviour of Mg-MOF-74 at the crystal surface. \emph{Chem. Sci.} \textbf{2025}, \emph{16}, 9462--9471\relax
\mciteBstWouldAddEndPuncttrue
\mciteSetBstMidEndSepPunct{\mcitedefaultmidpunct}
{\mcitedefaultendpunct}{\mcitedefaultseppunct}\relax
\EndOfBibitem
\bibitem[Mirzaei \latin{et~al.}(2026)Mirzaei, Mirzaei, Gao, Gándara, Wang, Tavani, Zhu, Proserpio, and Yaghi]{MOF3_f}
Mirzaei,~S.; Mirzaei,~M.~S.; Gao,~M.-Y.; Gándara,~F.; Wang,~K.; Tavani,~F.; Zhu,~H.; Proserpio,~D.~M.; Yaghi,~O.~M. Single-Crystalline Twelve-Connected Nanographene-Based Covalent Organic Frameworks. \emph{J. Am. Chem. Soc.} \textbf{2026}, \emph{148}, 25241--25248\relax
\mciteBstWouldAddEndPuncttrue
\mciteSetBstMidEndSepPunct{\mcitedefaultmidpunct}
{\mcitedefaultendpunct}{\mcitedefaultseppunct}\relax
\EndOfBibitem
\bibitem[Tavani \latin{et~al.}(2023)Tavani, Tofoni, and D’Angelo]{catal13101338}
Tavani,~F.; Tofoni,~A.; D’Angelo,~P. Exploring the Methane to Methanol Oxidation over Iron and Copper Sites in Metal–Organic Frameworks. \emph{Catalysts} \textbf{2023}, \emph{13}\relax
\mciteBstWouldAddEndPuncttrue
\mciteSetBstMidEndSepPunct{\mcitedefaultmidpunct}
{\mcitedefaultendpunct}{\mcitedefaultseppunct}\relax
\EndOfBibitem
\bibitem[Hofmann \latin{et~al.}(2026)Hofmann, Roslend, Ajello, Vornholt, Akinsoji, Hupp, and Chapman]{hofmann2026colossal}
Hofmann,~J.; Roslend,~A.; Ajello,~J.~G.; Vornholt,~S.~M.; Akinsoji,~O.; Hupp,~J.~T.; Chapman,~K.~W. Colossal Negative Thermal Expansion in MOF-808. \emph{J. Am. Chem. Soc.} \textbf{2026}, \emph{148}, 38404--38411\relax
\mciteBstWouldAddEndPuncttrue
\mciteSetBstMidEndSepPunct{\mcitedefaultmidpunct}
{\mcitedefaultendpunct}{\mcitedefaultseppunct}\relax
\EndOfBibitem
\end{mcitethebibliography}
\providecommand{\latin}[1]{#1}
\makeatletter
\providecommand{\doi}
  {\begingroup\let\do\@makeother\dospecials
  \catcode`\{=1 \catcode`\}=2 \doi@aux}
\providecommand{\doi@aux}[1]{\endgroup\texttt{#1}}
\makeatother
\providecommand*\mcitethebibliography{\thebibliography}
\csname @ifundefined\endcsname{endmcitethebibliography}  {\let\endmcitethebibliography\endthebibliography}{}

\end{document}